%% file: main.tex
\documentclass[runningheads]{llncs}

\usepackage{algpseudocode}
\usepackage{mathtools}
\usepackage{array}
\usepackage{stmaryrd}
\usepackage{amsmath}
\usepackage{mathtools}
\usepackage{amsfonts}
\usepackage{tikz}
\usepackage{pgfplots}
\usepackage{caption}
\usepackage{subcaption}
\usepackage{pdfpages}
\usepackage{wrapfig}
\usepackage{xcolor}
\usepackage[most]{tcolorbox}
\usepackage{listings}
\usepackage{xspace}
\usepackage{galois}

\usepackage{multirow}

\usepackage{apxproof}
\newtheoremrep{theorem}{Theorem}

\usepackage{hyperref}
\hypersetup{pdfcreator={}}

\definecolor{white}{rgb}{1,1,1}
\definecolor{mygreen}{rgb}{0,0.4,0}
\definecolor{light_gray}{rgb}{0.97,0.97,0.97}
\definecolor{mykey}{rgb}{0.117,0.403,0.713}

\tcbuselibrary{listings}
\newlength\inwd
\newcounter{ipythcntr}
\renewcommand{\theipythcntr}{\texttt{[\arabic{ipythcntr}]}}

\newtcblisting{pyin2}[1][]{%
  sharp corners,
  enlarge left by=\inwd,
  width=\linewidth-\inwd,
  enhanced,
  boxrule=0pt,
  colback=light_gray,
  listing only,
  top=0pt,
  bottom=0pt,
  overlay={
    \node[
      anchor=north east,
      text width=\inwd,
      font=\scriptsize\ttfamily\color{mykey},
      inner ysep=2mm,
      inner xsep=0pt,
      outer sep=0pt
      ] 
      at (frame.north west)
      {\refstepcounter{ipythcntr}\label{#1}In \theipythcntr:};
  }
  listing engine=listing,
  listing options={
    aboveskip=1pt,
    belowskip=1pt,
    basicstyle=\scriptsize\ttfamily,
    language=Python,
    keywordstyle=\color{mykey},
    showstringspaces=false,
    stringstyle=\color{mygreen},
    mathescape
  },
}

\newtcblisting{pyin}[1][]{%
  sharp corners,
  enlarge left by=\inwd,
  width=\linewidth-\inwd,
  enhanced,
  boxrule=0pt,
  colback=light_gray,
  listing only,
  top=0pt,
  bottom=0pt,
  overlay={
    \node[
      anchor=north east,
      text width=\inwd,
      font=\scriptsize\ttfamily\color{mykey},
      inner ysep=2mm,
      inner xsep=0pt,
      outer sep=0pt
      ] 
      at (frame.north west)
      {\refstepcounter{ipythcntr}\label{#1}In \theipythcntr:};
  }
  listing engine=listing,
  listing options={
    aboveskip=1pt,
    belowskip=1pt,
    basicstyle=\scriptsize\ttfamily,
    language=Python,
    keywordstyle=\color{mykey},
    showstringspaces=false,
    stringstyle=\color{mygreen},
    mathescape
  },
}
\newtcblisting{pyprint}{
  sharp corners,
  enlarge left by=\inwd,
  width=\linewidth-\inwd,
  enhanced,
  boxrule=0pt,
  colback=white,
  listing only,
  top=0pt,
  bottom=0pt,
  overlay={
    \node[
      anchor=north east,
      text width=\inwd,
      font=\footnotesize\ttfamily\color{mykey},
      inner ysep=2mm,
      inner xsep=0pt,
      outer sep=0pt
      ] 
      at (frame.north west)
      {};
  }
  listing engine=listing,
  listing options={
      aboveskip=1pt,
      belowskip=1pt,
      basicstyle=\footnotesize\ttfamily,
      language=Python,
      keywordstyle=\color{mykey},
      showstringspaces=false,
      stringstyle=\color{mygreen}
    },
}
\newtcblisting{pyout}[1][\theipythcntr]{
  sharp corners,
  enlarge left by=\inwd,
  width=\linewidth-\inwd,
  enhanced,
  boxrule=0pt,
  colback=white,
  listing only,
  top=0pt,
  bottom=0pt,
  overlay={
    \node[
      anchor=north east,
      text width=\inwd,
      font=\scriptsize\ttfamily\color{mykey},
      inner ysep=2mm,
      inner xsep=0pt,
      outer sep=0pt
      ] 
      at (frame.north west)
      {\setcounter{ipythcntr}{\value{ipythcntr}}Out#1:};
  }
  listing engine=listing,
  listing options={
      aboveskip=1pt,
      belowskip=1pt,
      basicstyle=\scriptsize\ttfamily,
      language=Python,
      keywordstyle=\color{mykey},
      showstringspaces=false,
      stringstyle=\color{mygreen}
    },
}

\newcommand{\header}[1]{\ensuremath\mbox{\textsf{hdr}}(#1)\xspace}
\newcommand{\idx}[1]{\ensuremath\mbox{\textsf{idx}}(#1)\xspace}
\newcommand{\values}{\ensuremath{\mathbb{V}}\xspace} 
\newcommand{\dataframes}{\ensuremath{\mathbb{D}}\xspace}
\newcommand{\defined}{\ensuremath{\stackrel{\mbox{\tiny def}}{=}}} 
\newcommand{\dom}{\ensuremath{{\scriptstyle\mathrm{dom}}}} 
\newcommand{\nat}{\ensuremath{\mathbb{N}}} 
\newcommand{\set}[1]{\ensuremath{\left\{#1\right\}}} 
\newcommand{\powerset}[1]{\ensuremath{\mathcal{P}\left(#1\right)}} 

\newcommand*{\lcdot}{\raisebox{-0.5ex}{\scalebox{1.5}{$\cdot$}}}

\def\raisefixb#1{
  \ifx#1\displaystyle
    \raise0.4ex
  \else
    \ifx#1\textstyle
      \raise0.4ex
    \else
      \ifx#1\scriptstyle
        \raise.180ex
      \else
        \raise.150ex
      \fi
    \fi
  \fi
} 
\newcommand{\bulcup}{\ooalign{$\cup$\cr\hidewidth\raise.225ex\hbox{$\lcdot\mkern2.25mu$}\cr}} 
\newcommand{\circcup}{\mathrel{\ooalign{$\cup$\cr\hidewidth\raise.1ex\hbox{$\circ\mkern1.5mu$}\cr}}} 
\newcommand{\bulsqcup}{\ooalign{$\sqcup$\cr\hidewidth\raise.225ex\hbox{$\lcdot\mkern2.25mu$}\cr}} 
\newcommand{\bulsqcap}{\ooalign{$\sqcap$\cr\hidewidth\raise.4ex\hbox{$\lcdot\mkern2.25mu$}\cr}} 
\newcommand{\final}{\ensuremath{\Omega}} 
\newcommand{\inputs}{\ensuremath{\mathrm{I}}} 
\newcommand{\maximal}{\ensuremath{^{+\infty}}} 
\newcommand{\operational}{\ensuremath{\langle \Sigma,\tau \rangle}} 
\newcommand{\outputs}{\ensuremath{\mathrm{U}}} 
\newcommand{\otrain}{\ensuremath{\outputs^{\mathrm{train}}}} 
\newcommand{\otest}{\ensuremath{\outputs^{\mathrm{test}}}} 
\newcommand{\tuple}[2]{\ensuremath{\langle #1, #2 \rangle}\xspace}
\newcommand{\vars}{\ensuremath{\mathbb{X}}\xspace} 

\newcommand{\pproperty}{\ensuremath{\mathcal{H}}\xspace}
\newcommand{\noleakage}{\ensuremath{\mathcal{I}}\xspace} 
\newcommand{\independent}{\textsc{ind}}
\newcommand{\unchanged}{\textsc{unch}}
\newcommand{\semantics}{\ensuremath{\powerset{\Sigma\maximal}}}
\newcommand{\properties}{\ensuremath{\powerset{\semantics}}}
\newcommand{\traceS}[1]{\ensuremath{\llbracket #1 \rrbracket}} %
\newcommand{\collectingS}[1]{\ensuremath{\llparenthesis #1 \rrparenthesis}} 

\newcommand{\dependency}{\rightsquigarrow}

\newcommand{\cdom}{\ensuremath{\mathbb{C}}\xspace}
\newcommand{\celm}{\ensuremath{\mathcal{C}}\xspace}
\newcommand{\rdom}{\ensuremath{\mathbb{R}}\xspace}
\newcommand{\relm}{\ensuremath{\mathcal{R}}\xspace}
\newcommand{\ddom}{\ensuremath{\mathbb{L}}\xspace}
\newcommand{\delm}{\ensuremath{\mathcal{L}}\xspace}

\newcommand{\sdom}{\ensuremath{\mathbb{S}}\xspace}
\newcommand{\selm}{\ensuremath{\mathcal{S}}\xspace}
\newcommand{\adom}{\ensuremath{\mathbb{A}}\xspace}
\newcommand{\aelm}{\ensuremath{\mathcal{A}}\xspace}

\allowdisplaybreaks[4]

\begin{document}


\title{An Abstract Interpretation-Based \\ Data 
Leakage Static Analysis}


\author{Filip Drobnjakovi\'{c}\inst{1}
\and 
Pavle Suboti\'{c}\inst{1}
\and
Caterina Urban\inst{2}
}
\institute{
Microsoft, Serbia
\and
Inria \& ENS $\mid$ PSL, France
}




\maketitle

\input{abstract}

%



\input{introduction.tex}

\input{background.tex}
\input{semantics.tex}

\input{technical.tex}
\input{evaluation.tex}
\input{related.tex}

\input{conclusion.tex}

%


\paragraph{Acknowledgements.}
We thank our colleagues at Microsoft Azure Data Labs and Microsoft Development Centre Serbia for all their feedback and support.

\bibliographystyle{splncs04}
\bibliography{main}

 \newpage

 \nosectionappendix
 \begin{toappendix}
 \input{appendix}
 \end{toappendix}

\end{document}

%% file: abstract.tex
\begin{abstract}
Data leakage is a well-known problem in machine learning which occurs 
when the training and testing datasets are not independent.
This phenomenon leads to 
overly optimistic accuracy estimates at training time, followed by 
a significant drop in performance when models are deployed in the real world. This can be dangerous, notably when models are used for risk prediction in high-stakes applications. 

In this paper, we propose an abstract interpretation-based static analysis to prove the absence of
data leakage. 
We implemented it 
in the \textsc{NBLyzer} 
framework and we demonstrate 
its performance and precision on $2111$ Jupyter 
notebooks from the Kaggle competition platform. 
\end{abstract}

%% file: introduction.tex
\section{Introduction}
As artificial intelligence (AI) continues its unprecedented impact on society, ensuring machine learning (ML)
models are accurate is crucial. 
To this end, ML models
need 
to be correctly trained and tested. This iterative task is typically performed 
within data science notebook environments~\cite{popularity,jetbrains}. 
A notable 
bug that can be introduced during this process is known as a \emph{data leakage}~\cite{dataleak}. Data leakages have been identified as a pervasive problem by the data science community~\cite{KapoorN23,Kaufman,Nisbet}.
In a number of recent cases data leakages crippled the performance of real-world risk prediction systems with dangerous consequences in high-stakes applications such as child welfare~\cite{ChouldechovaPFV18} and healthcare~\cite{Wong2021ExternalVO}.



Data 
leakages arise when dependent data is used to train and test a model. This can come in the form of 
overlapping data sets or, more insidiously, by library transformations that create indirect data dependencies. 

\begin{example} [Motivating Example] \label{ex:notebook}
Consider the following excerpt of a data science notebook (based on \emph{569.ipynb} from our benchmarks, and written in the small language that we introduce in Section~\ref{subsec:semantics}):
\begin{lstlisting}[
xleftmargin = 1cm,
framexleftmargin = 1em,
aboveskip=1.5em,
belowskip=1em,
language = Python,
basicstyle=\footnotesize\ttfamily,
stringstyle=\color{mygreen},
    keywordstyle=\color{mykey},
    showstringspaces=false,
                   mathescape,
                       breakatwhitespace=false,
    breaklines=true,
    captionpos=b,
    keepspaces=true,
    numbers=left,
    showspaces=false,
    showstringspaces=false,
    showtabs=false,
    tabsize=2]
data = read("data.csv")
X_norm = normalize(X)
X_train = X_norm.select[[$\lfloor 0.025 * R_{\texttt{X\_norm}} \rfloor + 1, \dots, R_{\texttt{X\_norm}}$]][]
X_test = X_norm.select[[0, $\dots$, $\lfloor 0.025 * R_{\texttt{X\_norm}} \rfloor$]][]
train(X_train)
test(X_test)
\end{lstlisting}
Line $1$ reads data from a CSV file and line $2$ normalizes it. Line $3$ and $4$ split the data into training and testing segments (we write $R_x$ for the number of data rows stored in $x$). Finally, line $5$ trains a ML model, and line $6$ tests its accuracy.









In this case, a data leakage is introduced because line $2$ performs a normalization, \emph{before} line $3$ and $4$ split into train and test data. This implicitly uses the mean over the entire dataset to perform the data transformation and, as a result, the train and test data are implicitly dependent on each other. In our experimental evaluation (cf. Section~\ref{sec:evaluation}) we found that this is a common pattern for introducing a data leakage in several real-world notebooks.
In the following, we say that the data resulting from the normalization done in line $2$ is \emph{tainted}.

\end{example}


Mainstream methods rely on detecting data leakages retroactively~\cite{Kaufman,dataleak}. Given a suspicious result, e.g., an overly accurate model, data analysis 
methods are used to identify data dependencies. 
However, 
a reasonable result may avoid suspicion from a data scientist 
until the model is already deployed. 
This is a natural use case for  \emph{static analysis} to detect data leakages at development time.

In this paper, we propose a static analysis for proving the absence of data leakage in data-manipulating programs: it tracks the origin of data used for training and testing and verifies that they originate from \emph{disjoint} and \emph{untainted} data sources. 
In Example~\ref{ex:notebook} our analysis identifies a data leakage since \texttt{X\_train} and \texttt{X\_test} originate from previously normalized data (despite being disjoint).

Our static analysis (cf. Section~\ref{sec:analysis}) is designed 
within the abstract interpretation framework~\cite{CC77}: it is derived through successive abstractions from the (sound and complete, but not computable) collecting program semantics (cf. Section~\ref{sec:semantics}). This formal development allows us to formally justify the soundness of the analysis (cf. Theorem~\ref{thm:soundness}), and to exactly pinpoint where it can lose precision (e.g., modeling data joins, cf. Section~\ref{subsec:analysis}) to guide the design of more precise abstractions, if necessary in the future (in our evaluation we found the current analysis to be sufficiently precise, cf. Section~\ref{sec:evaluation}).
Moreover, it allows a clear comparisons with other related static analyses, e.g., information flow and taint analyses (cf. Section~\ref{sec:rel}).
Finally, this design principle allowed us to identify and overcome issues and shortcomings of previous data leakage analysis attempts~\cite{NBLyzer,SuboticBS22}.

We implemented our analysis 
in the \textsc{NBLyzer}~\cite{NBLyzer} framework. 
We evaluate its performance on $2111$ Jupyter notebooks from the Kaggle competition platform, and demonstrate that our approach scales to the performance constraints of interactive data science notebook environments while detecting $25$ real data leakages with a precision of 93\%. Notably, we are able to detect 60\% more data leakages compared to the ad-hoc analysis previously implemented in \textsc{NBLyzer}.

%% file: background.tex
\section{Background}
\label{sec:background}

\subsection{Data Frame-Manipulating Programs}\label{sec:dataframes}

We consider programs manipulating data frames, that is, tabular data structures with columns labeled by non-empty unique names. Let $\values$ be a set of (heterogeneous atomic) values (i.e., such as numerical or string values). We can formalize a data frame as a possibly empty $(r \times c)$-matrix of values, where $r \in \nat$ and $c \in \nat$ denote the number of matrix rows and columns, respectively. The first row of non-empty data frames contains the labels of the data frame columns. Let 
$    \dataframes \defined \bigcup_{r \in \nat} \bigcup_{c \in \nat} \values^{r \times c}$
be the set of all possible data frame. Given a data frame $D \in \dataframes$, we use $R_D$ and $C_D$ to denote the number of its rows and columns, respectively, and write $\header{D}$ for the set of labels of its columns.
We also write $D[r]$ for the specific row indexed with $r \in R_D$ in $D$.

\subsection{Trace Semantics}\label{sec:traces}

The \emph{semantics} of a data frame-manipulating program is a mathematical characterization of its behavior when executed for all 
possible input data. 
We model the operational 
semantics of a program
in a language-independent way
as a \emph{transition system} $\operational$, where $\Sigma$ is a (potentially infinite) 
set of program states and the transition relation $\tau \subseteq \Sigma \times \Sigma$ describes the possible 
transitions between states.
The set of \emph{final states} of the program is $\final \defined \{s \in \Sigma~|~\forall s' \in \Sigma:~\langle s,s' \rangle \not\in \tau\}$.

In the following, let 
$\Sigma\maximal \defined \Sigma^+ \mathrel{\cup} \Sigma^\omega$ 
be the set of all non-empty finite or infinite sequences of program states. 
%
A \emph{trace} is a non-empty sequence of program states that respects the transition relation $\tau$, that is, $\langle s,s' \rangle \in \tau$ for each pair of consecutive states 
$s, s' \in \Sigma$ in the sequence. 
The \emph{trace semantics} 
$\Upsilon \in \powerset{\Sigma\maximal}$ 
generated by a transition system $\operational$ is the union of all finite traces that 
are terminating with a final state in $\final$, and all infinite traces \cite{Cousot02}:
\begin{equation}\label{eq:maximal}
\begin{aligned}
\Upsilon \defined &\bigcup_{n \in \nat^+}
\set{	
	s_0 \dots s_{n-1} \in \Sigma^n \mid \forall i < n-1\colon \tuple{s_i}{s_{i+1}} \in \tau, 
	s_{n-1} \in 
	\final} \\ 
&\cup~
\set{	
	s_0 \dots \in \Sigma^\omega \mid \forall i \in \nat\colon \tuple{s_i}{s_{i+1}} \in \tau
} 
\end{aligned}
\end{equation}
In the rest of the paper, we write $\traceS{P}$ for 
the trace semantics of a program $P$.


%% file: semantics.tex

\section{Concrete Data Leakage Semantics}\label{sec:semantics}

The trace semantics 
fully describes the behavior of a program. 
However, reasoning about a 
particular property of a program 
is 
facilitated by the design of a semantics that abstracts away from irrelevant details about program executions. In 
this section, we define our property of interest --- absence of data leakage --- and 
use abstract interpretation 
to 
systematically derive, by 
abstraction of the trace semantics, a semantics that precisely captures this property.

\subsection{(Absence of) Data Leakage}\label{subsec:property}
\looseness=-1
We use an extensional definition of a \emph{property} as the set of elements having such a property \cite{CC77,aiproduct}. 
This allows checking property satisfaction by set inclusion (see below) also across abstractions (cf. Theorems~\ref{thm:noleakage-dependency} and~\ref{thm:noleakage-semantics}). 
Semantic properties of programs are properties of their semantics.  
Thus, properties of programs with trace semantics in $\semantics$ are sets of sets of traces
in $\properties$. 
The set of program properties
forms a complete boolean lattice $\langle \properties, \subseteq,
\cup, \cap, \emptyset, \semantics \rangle$ for subset inclusion (i.e., logical implication). 
The strongest property is the
standard 
\emph{collecting semantics} $\Lambda \in \properties$:
$\Lambda \defined \set{\Upsilon}$,
i.e., 
the property of ``being the program with trace semantics $\Upsilon$''.
Let $\collectingS{P}$ denote the collecting
semantics of a program
$P$. Then, a program $P$ satisfies a
given property $\pproperty \in \properties$ if and only if its
collecting semantics is a subset of
$\pproperty$:
$P \models \pproperty
\Leftrightarrow
\collectingS{P} \subseteq \pproperty$.
%
In this paper, we consider the property of \emph{absence of data leakage}, which requires data used for training and data used for testing a machine learning model to be \emph{independent}.

\begin{example}[(In)dependent Data Frame Variables]\label{ex:independent}
    Let us consider a program $P$ with a single input data frame variable reading data frames with four rows and one single column with values in $\set{3, 9}$, i.e., data frames in $\bigcup_{r \in \set{1, 2, 3, 4}} \set{3, 9}^r$ (cf. Section~\ref{sec:dataframes}). Imagine that $P$ first performs min-max normalization (i.e., rescaling all data frame values to be in the $[0, 1]$ range) and then splits the data frame in half to use the first two rows for training and the last two rows for testing. The table below shows all possible train and test data resulting from all possible input data of this program:
\begin{center}
    \begin{tabular}{c|c|c|c|c|c|c|c|c|c|c|c|c|c|c|c|c|r}
   \multirow{4}{*}{\textbf{input data}} & 3 & 3 & 3 & 3 & 9 & 9 & 9 & 9 & 3 & 3 & 3 & 3 & 9 & 9 & 9 & 9 & \quad\textcolor{gray}{1}\\
    & 3 & 3 & 9 & 9 & 3 & 3 & 9 & 9 & 3 & 3 & 9 & 9 & 3 & 3 & 9 & 9 & \quad\textcolor{gray}{2}\\
    & 3 & 9 & 3 & 9 & 3 & 9 & 3 & 9 & 3 & 9 & 3 & 9 & 3 & 9 & 3 & 9 & \quad\textcolor{gray}{3} \\
    & 3 & 3 & 3 & 3 & 3 & 3 & 3 & 3 & 9 & 9 & 9 & 9 & 9 & 9 & 9 & 9 & \quad\textcolor{gray}{4} \\
    & $\downarrow$ & $\downarrow$ & $\downarrow$ & $\downarrow$ & $\downarrow$ & $\downarrow$ & $\downarrow$ & $\downarrow$ & $\downarrow$ & $\downarrow$ & $\downarrow$ & $\downarrow$ & $\downarrow$ & $\downarrow$ & $\downarrow$ & $\downarrow$ \\
    \multirow{2}{*}{\textbf{train data}} & 0 & 0 & 0 & 0 & 1 & 1 & 1 & 1 & 0 & 0 & 0 & 0 & 1 & 1 & 1 & 0 & \quad\textcolor{gray}{1} \\
    & 0 & 0 & 1 & 1 & 0 & 0 & 1 & 1 & 0 & 0 & 1 & 1 & 0 & 0 & 1 & 0 & \quad\textcolor{gray}{2} \\
    \cline{0-16}
    \multirow{2}{*}{\textbf{test data}} & 0 & 1 & 0 & 1 & 0 & 1 & 0 & 1 & 0 & 1 & 0 & 1 & 0 & 1 & 0 & 0 & \quad\textcolor{gray}{1} \\
    & 0 & 0 & 0 & 0 & 0 & 0 & 0 & 0 & 1 & 1 & 1 & 1 & 1 & 1 & 1 & 0 & \quad\textcolor{gray}{2} \\
    \multicolumn{12}{c}{} & \multicolumn{1}{c}{\textcolor{gray}{$\sigma$}} & \multicolumn{3}{c}{} & \multicolumn{1}{c}{\textcolor{gray}{$\sigma'$}}
    \end{tabular}
\end{center}
\looseness=-1
In this case, train and test data are \emph{not} independent: if we consider, for instance, the execution $\sigma$ with input data frame value ``3$\mid$9$\mid$9$\mid$9'' we can change the value of its first row (i.e., $r = 1$ in Equation~\ref{eq:independent}) from $\bar{v} = 3$ to $\bar{v} = 9$ (while leaving all other rows unchanged) to obtain an execution $\sigma'$ resulting in a difference in \emph{both} train and test data (i.e., $\sigma(\otrain_P) \not= \sigma'(\otrain_P)$ and $\sigma(\otest_P) \not= \sigma'(\otest_P)$ in Equation~\ref{eq:independent}, with train data differing at line $2$ and test data differing at both lines $1$ and $2$).

Instead, the table below shows all possible resulting train and test data 
if the 
normalization is performed \emph{after} the split into train and test data:
\begin{center}
    \begin{tabular}{c|c|c|c|c|c|c|c|c|c|c|c|c|c|c|c|c|r}
   \multirow{4}{*}{\textbf{input data}} & 3 & 3 & 3 & 3 & 9 & 9 & 9 & 9 & 3 & 3 & 3 & 3 & 9 & 9 & 9 & 9 & \quad\textcolor{gray}{1} \\
    & 3 & 3 & 9 & 9 & 3 & 3 & 9 & 9 & 3 & 3 & 9 & 9 & 3 & 3 & 9 & 9 & \quad\textcolor{gray}{2} \\
    & 3 & 9 & 3 & 9 & 3 & 9 & 3 & 9 & 3 & 9 & 3 & 9 & 3 & 9 & 3 & 9 & \quad\textcolor{gray}{3} \\
    & 3 & 3 & 3 & 3 & 3 & 3 & 3 & 3 & 9 & 9 & 9 & 9 & 9 & 9 & 9 & 9 & \quad\textcolor{gray}{4} \\
    & $\downarrow$ & $\downarrow$ & $\downarrow$ & $\downarrow$ & $\downarrow$ & $\downarrow$ & $\downarrow$ & $\downarrow$ & $\downarrow$ & $\downarrow$ & $\downarrow$ & $\downarrow$ & $\downarrow$ & $\downarrow$ & $\downarrow$ & $\downarrow$ \\
    \multirow{2}{*}{\textbf{train data}} & 0 & 0 & 0 & 0 & 1 & 1 & 0 & 0 & 0 & 0 & 0 & 0 & 1 & 1 & 0 & 0 & \quad\textcolor{gray}{1} \\
    & 0 & 0 & 1 & 1 & 0 & 0 & 0 & 0 & 0 & 0 & 1 & 1 & 0 & 0 & 0 & 0 & \quad\textcolor{gray}{2} \\
    \cline{0-16}
    \multirow{2}{*}{\textbf{test data}} & 0 & 1 & 0 & 1 & 0 & 1 & 0 & 1 & 0 & 0 & 0 & 0 & 0 & 0 & 0 & 0 & \quad\textcolor{gray}{1} \\
    & 0 & 0 & 0 & 0 & 0 & 0 & 0 & 0 & 1 & 0 & 1 & 0 & 1 & 0 & 1 & 0 & \quad\textcolor{gray}{2} \\
    \end{tabular}
\end{center}
Here train and test data remain independent as modifying any input data row $r$ in any execution 
yields another execution 
that may result in a difference in either train and test data \emph{but never both}. Equivalently, all possible values of \emph{either} train and test data are possible independently of the choice of the value of the row $r$.
\end{example}

\looseness=-1
More formally, let $\vars$ be the set of all the (data frame) variables of a (data frame-manipulating) program $P$. We denote with $\inputs_P \subseteq \vars$ the set of its \emph{input} or source data frame variables, i.e., data frame variables whose value is directly read from the input, and use $\outputs_P \subseteq \vars$ to denote the set of its \emph{used} data frame variables, i.e., data frame variables used for training or testing a ML model. We write $\otrain_P \subseteq \outputs_P$ and $\otest_P \subseteq \outputs_P$ for the variables used for training and testing, respectively. 
For simplicity, we can assume that programs are in static single-assignment form so that data frame variables are assigned exactly once: data is read from the input, transformed and normalized, and ultimately used for training and testing. 
Given a trace $\sigma \in \traceS{P}$, we 
write $\sigma(i)$ and $\sigma(o)$ to denote the value of the data frame variables $i \in \inputs_P$ and $o \in \outputs_P$ in $\sigma$. We can now define when used data frame variables are independent in a program with trace semantics $\traceS{P}$:
\begin{equation}\label{eq:independent}
\begin{aligned}
&\independent(\traceS{P})~\defined~
\forall \sigma \in \traceS{P}, i \in \inputs_P, r \in R_i \colon \unchanged(\sigma, i, r, \otest_P)   
\lor \unchanged(\sigma, i, r, \otrain_P)  \\
&\unchanged(\sigma, i, r, U)~\defined~ 
\forall \bar{v} \in \values^{C_i}\colon 
\sigma(i)[r] \not= \bar{v} 
\Rightarrow \left( \exists \sigma' \in \traceS{P}\colon \right. \\ 
&\qquad 
 \sigma'(i)[r] \!= \bar{v} \land \sigma(i) \stackrel{\bar{r}}{=} \sigma'(i) \land 
 \sigma(\inputs_P \!\setminus\! \set{i}) = \sigma'(\inputs_P \!\setminus\! \set{i}) \land
 \sigma(U) = \sigma'(U) )
\end{aligned}
\end{equation}
where $R_i$ and $C_i$ stand for $R_{\sigma(i)}$ (i.e., number of rows of the data frame value of $i \in \inputs_P$)
and $C_{\sigma(i)}$ (i.e., number of columns of the data frame value of $i \in \inputs_P$), respectively, 
$\sigma(i) \stackrel{\bar{r}}{=} \sigma'(i)$ stands for $\forall r' \in R_i\colon r' \not= r \Rightarrow \sigma(i)[r'] = \sigma'(i)[r']$ (i.e., the data frame value of $i \in \inputs_P$ remains unchanged for any row $r' \not= r$), and $\sigma(X) = \sigma'(X)$ stands for $\forall x \in X\colon \sigma(x) = \sigma'(x)$.
The definition requires that changing the value of a data source $i \in \inputs_P$ can modify data frame variables used for training ($\otrain_P$) or testing ($\otest_P$), \emph{but not both}: the value of data frame variables used for either training or testing in a trace $\sigma$ remains the same independently of all possible values $\bar{v} \in \values^{C_i}$ of any portion (e.g., any row $r \in R_i$) of any input data frame variable $i \in \inputs_P$ in $\sigma$. 
Note that this definition quantifies over changes in data frame rows since the split into train and test data happens across rows (e.g., using \verb+train_test_split+ in Pandas), but takes into account all possible column values in each row ($\bar{v} \in \values^{C_i}$). It also implicitly takes into account implicit flows of information by considering traces in $\traceS{P}$. In particular, 
in terms of secure information flow, notably non-interference, this definition says that \emph{we cannot simultaneously observe different values} in $\otrain_P$ and $\otest_P$, regardless of the values of the input data frame variables. Here we weaken non-interference to consider either $\otrain_P$ or $\otest_P$ as low outputs (depending on which row of the input data frame variables is modified), instead of fixing the choice beforehand.
Note also that $\unchanged$ quantifies over all possible values $\bar{v} \in \values^{C_i}$ of the changed row $i$ rather than quantifying over traces to allow non-determinism, i.e., not all traces that only differ at row $r \in R_i$ of data frame variable $i \in \inputs_P$ need to agree on the values of the used variables $U \subseteq U_P$ but all values of $U$ that are feasible from a value of $r$ of $i$ need to be feasible for \emph{all} possible values of $r$ of $i$.
In terms of input data (non-)usage \cite{input}, this definition says that training and testing \emph{do not use} the same (portions of the) input data sources. Here we generalize the notion of data usage proposed by Urban and Müller \cite{input} to multi-dimensional variables and allow multiple values for all outcomes but one (variables used for either training or testing) for each variation in the values of the input variables.

The absence of data leakage property 
can now be formally defined as the set 
    $\noleakage \defined \set{ \traceS{P} \in \semantics \mid \independent(\traceS{P})}$ 
of programs 
(semantics) that use independent data for training and testing ML models. Thus 
$P \models \noleakage \Leftrightarrow \collectingS{P} \subseteq \noleakage$.


In the rest of this section, we 
derive, by abstraction of the collecting semantics $\Lambda$, a \emph{sound} and \emph{complete} semantics $\dot{\Lambda}_I$ that contains only and exactly the information needed to reason about (the absence of) data leakage. A further abstraction in the next section, loses completeness but yields a \emph{sound} and \emph{computable} over-approximation of $\dot{\Lambda}_I$ that allows designing a static analysis to effectively detect data leakage in data frame-manipulating programs.

\subsection{Dependency Semantics}

From the definition of absence of data leakage, 
we observe that for reasoning about data leakage we essentially need to track the flow of information between (portions of) input data sources and data used for training or testing. Thus we can abstract the collecting semantics into a set of dependencies between (rows of) input data frame variables and used data frame variables.  

We define the following Galois connection:
\begin{equation}\label{eq:dependency0}
\tuple{\properties}{\subseteq}
	\galois{\alpha_{I \dependency U}}{\gamma_{I \dependency U}}
\tuple{\powerset{(\vars \times \nat) \times (\vars \times \nat)}}{\supseteq}
\end{equation}
\looseness=-1
between sets of sets of traces and sets of relations (i.e., dependencies) between data frame variables indexed at some row. The abstraction and concretization function are parameterized by a set $I \subseteq \vars$ of input variables and a set $U \subseteq \vars$ of used variables of interest. In particular, 
the dependency abstraction $\alpha_{I \dependency U}\colon \properties \rightarrow \powerset{(\vars \times \nat) \times (\vars \times \nat)}$ is:
\begin{equation*}\label{eq:dependency1}
\alpha_{I \dependency U}(S) \defined \set{
i[r] \dependency o[r'] ~\middle\vert~
\begin{matrix}
i \in I, r \in \nat, o \in U, r' \in \nat, (\forall T \in S\colon \\
\exists \sigma \in T, \bar{v} \in \values^{C_i} \colon 
\forall \sigma' \in T\colon \sigma(i) \stackrel{\bar{r}}{=} \sigma'(i)~\land \\
 \sigma(I \!\setminus\! \set{i}) = \sigma'(I \!\setminus\! \set{i}) \land
\sigma(o)[r'] = \sigma'(o)[r'] 
\\ \Rightarrow \sigma'(i)[r] \not= \bar{v})
\end{matrix}
}
\end{equation*}
where we write $i[r] \dependency o[r']$ for a dependency $\langle \langle i, r \rangle, \langle o, r' \rangle \rangle$ between a data frame variable $i \in I$ at the row indexed by $r \in \nat$ and a data frame variable $o \in U$ at the row indexed by $r' \in \nat$. In particular, $\alpha_{\dependency}$ extracts a dependency $i[r] \dependency o[r']$ when (in all sets of traces $T$ in the semantic property $S$) there is a value $\bar{v} \in \values^{C_i}$ for row $r$ of data frame variable $i$ that changes the value at row $r'$ of data frame variable $o$, that is, there is a value for row $r'$ of data frame variable $o$ that cannot be reached if the value for row $r$ of $i$ is changed to $\bar{v}$ (and all else remains the same, i.e., $\sigma(i) \stackrel{\bar{r}}{=} \sigma'(i) \land 
 \sigma(I \!\setminus\! \set{i}) = \sigma'(I \!\setminus\! \set{i})$). 

Note that our 
dependency abstraction generalizes that of Cousot~\cite{Cousot19} to non-deterministic programs and multi-dimensional data frame variables, thus tracking dependencies between portions of data frames. As in \cite{Cousot19}, this is an abstraction of semantic properties thus the dependencies must hold for all semantics having the semantic property: more semantics have a semantic property, fewer dependencies will hold for all semantics. Therefore, sets of dependencies are ordered by superset inclusion $\supseteq$ (cf. Equation~\ref{eq:dependency0}).

\begin{example}[Dependencies Between Data Frame Variables]\label{ex:dependency}
Let us consider again the program $P$ from Example~\ref{ex:independent}. 
Let $i$ denote the input data frame 
of the program and let $o_{\mathrm{train}}$ and $o_{\mathrm{test}}$ denote the data frames 
used for training and testing. 
In this case, for instance, 
we have 
$i[1] \dependency o_{\mathrm{train}}[2]$ because, taking execution $\sigma$, changing only the value of $i[1]$ from $3$ to $9$ yields execution $\sigma'$ which changes the value of $o_{\mathrm{train}}[2]$, i.e., all other executions either differ at other rows of $i$ or differ at least in the value of $o_{\mathrm{train}}[2]$ (such as $\sigma'$). 
In fact, the set of dependencies for the whole set of executions of the program shows that $o_{\mathrm{train}}$ and $o_{\mathrm{test}}$ depend on \emph{all} rows of the input data frame variable $i$.

Instead, performing 
normalization \emph{after} splitting into 
train / test data yields 
$\set{ i[1] \dependency o_{\mathrm{train}}[j], i[2] \dependency o_{\mathrm{train}}[j], i[3] \dependency o_{\mathrm{test}}[j], i[4] \dependency o_{\mathrm{test}}[j]}$, $j \in \set{1, 2}$, where $o_{\mathrm{train}}$ and $o_{\mathrm{test}}$ depend on disjoint subsets of rows of the input data frame 
$i$.
\end{example}

It is easy to see that the abstraction function $\alpha_{\inputs_P \dependency \outputs_P}$ is a complete join morphism. Thus, $\gamma_{\inputs_P \dependency \outputs_P}(D) \defined \bigcup \set{S \mid \alpha_{\inputs_P \dependency \outputs_P}(S) \supseteq D}$.

We can now define the \emph{dependency semantics} $\Lambda_{I \dependency U} \in \powerset{(\vars \times \nat) \times (\vars \times \nat)}$ by abstraction of the collecting semantics $\Lambda$:
    $\Lambda_{I \dependency U} \defined \alpha_{I \dependency U}(\Lambda)$.
In the rest of the paper, we
write $\collectingS{P}_{\dependency}$ to denote the dependency semantics of a program $P$, leaving the sets of data frame variables of interest $I$ and $U$ implicitly set to $\inputs_P$ and $\outputs_P$, respectively.
The dependency semantics remains sound and complete: 
\begin{theoremrep}
\label{thm:noleakage-dependency}
$P \models \noleakage \Leftrightarrow \collectingS{P}_{\dependency} \supseteq \alpha_{\inputs_P \dependency \outputs_P}(\noleakage)$
\end{theoremrep}
 \begin{appendixproof}
 Let $P \models \noleakage$. From the subset inclusion in Section~\ref{subsec:property}, we have that $\collectingS{P} \subseteq \noleakage$. Thus, from the Galois connection in Equation~\ref{eq:dependency0} (note the inverse $\supseteq$ order in the abstract domain!), we have $\alpha_{\inputs_P \dependency \outputs_P}(\collectingS{P}) \supseteq \alpha_{\inputs_P \dependency \outputs_P}(\noleakage)$. From the definition of $\collectingS{P}_{\dependency}$, 
 we can then conclude that $\collectingS{P}_{\dependency} \supseteq \alpha_{\inputs_P \dependency \outputs_P}(\noleakage)$.

 Vice versa, let $\collectingS{P}_{\dependency} \supseteq \alpha_{\inputs_P \dependency \outputs_P}(\noleakage)$. From the definition of $\collectingS{P}_{\dependency}$, we have $\alpha_{\inputs_P \dependency \outputs_P}(\collectingS{P}) \supseteq \alpha_{\inputs_P \dependency \outputs_P}(\noleakage)$, and from the Galois connection in Equation~\ref{eq:dependency0} we have $\collectingS{P} \subseteq \gamma_{\inputs_P \dependency \outputs_P}(\alpha_{\inputs_P \dependency \outputs_P}(\noleakage))$. From the definition of $\gamma_{\inputs_P \dependency \outputs_P}$, 
 we have $\collectingS{P} \subseteq \noleakage$ and we can thus conclude $\collectingS{P} \subseteq \noleakage$.
 \end{appendixproof}


\subsection{Data Leakage Semantics}\label{subsec:semantics}

\looseness=-1
As hinted by Example~\ref{ex:dependency}, we observe that for detecting data leakage (resp. verifying absence of data leakage), we care in particular about which rows of input data frame variables the used data frame variables depend on.
In case of data leakage (resp. absence of data leakage), data frame variables used for different purposes will depend on \emph{overlapping} (resp. \emph{disjoint}) sets of rows of input data frame variables.
Thus, we further abstract the dependency semantics $\Lambda_{\dependency^+}$ pointwise \cite{CousotC94} into a map for each data frame variable associating with each data frame row index the set of (input) 
variables (indexed at some row) from which it depends on.

Formally, we define the following Galois connection:
\begin{equation}\label{eq:semantics0}
\tuple{\powerset{(\vars \times \nat) \times (\vars \times \nat)}}{\supseteq}
	\galois{\dot{\alpha}}{\dot{\gamma}}
\tuple{\vars \rightarrow (\nat \rightarrow \powerset{\vars \times \nat})}{\dot{\supseteq}}
\end{equation}
where the abstraction 
$\dot{\alpha}\colon \powerset{(\vars \times \nat) \times (\vars \times \nat)} \rightarrow (\vars \rightarrow (\nat \rightarrow \powerset{\vars \times \nat}))$ is:
\begin{equation}\label{eq:semantics1}
    \dot{\alpha}(D) \defined \lambda x \in \vars\colon (
    \lambda r \in\nat\colon \set{i[r'] \mid i \in \vars, r' \in \nat, i[r'] \dependency x[r] \in D} 
    )
\end{equation}

\begin{example}[Data Leakage Semantics]\label{ex:semantics}
Let us consider again the last 
dependencies in Example~\ref{ex:dependency}: $\set{ i[1] \dependency o_{\mathrm{train}}[j], i[2] \dependency o_{\mathrm{train}}[j], i[3] \dependency o_{\mathrm{test}}[j], i[4] \dependency o_{\mathrm{test}}[j] }$, $j \in \set{1, 2}$.
Its abstraction following Equation~\ref{eq:semantics1} is the following map:
\begin{equation*}
    \lambda x\colon \begin{cases} \lambda r\colon \begin{cases}
        \set{ i[1], i[2] } & r = 1 \\
        \set{ i[1], i[2] } & r = 2 
    \end{cases} & x = o_{\mathrm{train}} \\
    \lambda r\colon \begin{cases}
        \set{ i[3], i[4] } & r = 1 \\
        \set{ i[3], i[4] } & r = 2 
    \end{cases} & x = o_{\mathrm{test}}
    \end{cases}
\end{equation*}

Instead, the abstraction of the set of dependencies resulting from performing 
normalization \emph{before} splitting into train and test data is the following map:
\begin{equation*}
    \lambda x\colon \begin{cases} \lambda r\colon \begin{cases}
        \set{ i[1], i[2], i[3], i[4] } & r = 1 \\
        \set{ i[1], i[2], i[3], i[4] } & r = 2 
    \end{cases} & x = o_{\mathrm{train}} \\
    \lambda r\colon \begin{cases}
        \set{ i[1], i[2], i[3], i[4] } & r = 1 \\
        \set{ i[1], i[2], i[3], i[4] } & r = 2 
    \end{cases} & x = o_{\mathrm{test}}
    \end{cases}
\end{equation*}
\end{example}

The abstraction function $\dot{\alpha}$ is another complete join morphism so it uniquely determines the concretization function: $\dot{\gamma}(m) \defined \bigcap \set{D \mid \dot{\alpha}(D) ~\dot{\supseteq}~ m}$.

We finally derive our \emph{data leakage semantics} $\dot{\Lambda} \in \vars \rightarrow (\nat \rightarrow \powerset{\vars \times \nat})$ by abstraction of the dependency semantics $\Lambda_{\dependency}$:
    $\dot{\Lambda} \defined \dot{\alpha}(\Lambda_{\dependency})$.
In the following, we write $\dot{\collectingS{P}}$ for the data leakage semantics of a program $P$.
The abstraction $\dot{\alpha}$ does not lose any information, so we still have both soundness and completeness: 
\begin{theoremrep}
\label{thm:noleakage-semantics}
$P \models \noleakage \Leftrightarrow \dot{\collectingS{P}} ~\dot{\supseteq}~ \dot{\alpha}(\alpha_{\inputs_P \dependency \outputs_P}(\noleakage))$
\end{theoremrep}
 \begin{appendixproof}
 The proof is analogous to that of Theorem~\ref{thm:noleakage-dependency}.
 Let $P \models \noleakage$. We have that $\collectingS{P}_{\dependency} \supseteq \alpha_{\inputs_P \dependency \outputs_P}(\noleakage)$ from Theorem~\ref{thm:noleakage-dependency}. From the Galois connection in Equation~\ref{eq:semantics0}, we have $\dot{\alpha}(\collectingS{P}_{\dependency}) ~\dot{\supseteq}~ \dot{\alpha}(\alpha_{\inputs_P \dependency \outputs_P}(\noleakage))$. Thus, from the definition of $\dot{\collectingS{P}}$, 
 we can conclude that $\dot{\collectingS{P}} ~\dot{\supseteq}~ \dot{\alpha}(\alpha_{\inputs_P \dependency \outputs_P}(\noleakage))$.

 Vice versa, let $\dot{\collectingS{P}} ~\dot{\supseteq}~ \dot{\alpha}(\alpha_{\inputs_P \dependency \outputs_P}(\noleakage))$. From the definition of $\dot{\collectingS{P}}$ we have $\dot{\alpha}(\collectingS{P}_{\dependency}) ~\dot{\supseteq}~ \dot{\alpha}(\alpha_{\inputs_P \dependency \outputs_P}(\noleakage))$, and from the Galois connection in Equation~\ref{eq:semantics0} we have $\collectingS{P}_{\dependency} ~\supseteq~ \dot{\gamma}(\dot{\alpha}(\alpha_{\inputs_P \dependency \outputs_P}(\noleakage)))$. From the definition of $\dot{\gamma}$ derived from $\dot{\alpha}$,
 we have $\collectingS{P}_{\dependency} ~\supseteq~ \alpha_{\inputs_P \dependency \outputs_P}(\noleakage)$ and from Theorem~\ref{thm:noleakage-dependency} we can thus conclude $P \models \noleakage$.
 \end{appendixproof}

We can now equivalently verify absence of data leakage by checking that data frames 
used for different purposes depend on disjoint (rows of) input data: 
\begin{lemma}\label{lm:check}
$P \models \noleakage \Leftrightarrow \forall o_1 \in \otrain_P, o_2 \in \otest_P\colon 
\dot{\collectingS{P}}o_1 
~\cap~ 
\dot{\collectingS{P}}o_2 = \emptyset$
where, with a slightly abuse of notation, $\dot{\collectingS{P}}o$ stands for $\bigcup_{r \in \dom(\dot{\collectingS{P}}o)} \dot{\collectingS{P}}o(r) $, i.e., the union of all sets $\dot{\collectingS{P}}o(r)$ of rows of input data frame variables in the range of $\dot{\collectingS{P}}o$ the data leakage semantics $\dot{\collectingS{P}}$ for the used data frame variable $o$
\end{lemma}

\begin{example}[Continued from Example~\ref{ex:semantics}]
    The first map in Example~\ref{ex:semantics} satisfies Lemma~\ref{lm:check} since the set $\set{ i[1], i[2] }$ of input data frame rows on which $o_{\mathrm{train}}$ depends is \emph{disjoint} from the set $\set{ i[3], i[4] }$ of input data frame rows on which $o_{\mathrm{test}}$ depends. Thus, performing min-max normalization \emph{after} splitting into train and test data does not create data leakage. 
    
    This is not the case for the second map in Example~\ref{ex:semantics} where the sets of input data frame rows from which $o_{\mathrm{train}}$ and $o_{\mathrm{test}}$ depend are identical, indicating data leakage when 
    normalization is done \emph{before} the split into train and test data.
\end{example}

\subsubsection{Small Data Frame-Manipulating Language}

\looseness=-1
The formal treatment 
so far is language independent. In the rest of this section, we give a constructive definition of our data leakage semantics $\dot{\Lambda}_I$ for a small data frame-manipulating language which we then use to illustrate our data leakage analysis in the next section. (Note that the actual implementation of the analysis handles more advanced constructs such as branches, loops, and procedures calls, cf. Appendix~\ref{sec:integration}).

We consider a simple sequential language without procedures nor references. The only variable data type is the set $\dataframes$ of data frames. 
Programs in the language are sequences of statements, which belong to either of the following classes:
\begin{enumerate}
\item \textbf{source}: $y = \text{read}(name) \quad\quad\quad\qquad \hspace{0.2em} name \in \mathbb{W}$ 

\item \textbf{select}: $y = x.\text{select}[\bar{r}][C] \quad\quad\quad\qquad \hspace{0.2em} \bar{r} \in \nat^{k \leq R_x}, C \subseteq \header{x}$

\item \textbf{merge}: $y = op(x_1, x_2) \qquad\qquad\qquad \hspace{0.2em} x_1, x_2 \in \vars, op \in \set{ \mbox{concat}, \mbox{join}}$ 


\item \textbf{function}: $y = f(x) \qquad\qquad\qquad\quad \hspace{0.4em} x \in \vars, f \in \{ \mbox{normalize}, \mbox{other} \}$

\item \textbf{use}: $f(X) \qquad\qquad\qquad\qquad\qquad\quad \hspace{0.3em} X \subseteq \vars, f \in \{ \mbox{train}, \mbox{test} \}$
\end{enumerate}
where $name \in \mathbb{W}$ is a (string) data file name; we 
write $R_x$ and $\header{x}$ for the number of rows and set of labels of the columns of the data frame (value) stored into the variable $x$. The \emph{source} statement (representing library functions such as \verb+read_csv+, \verb+read_excel+, etc., in Python \textrm{pandas}) reads data 
from an input file and stores it into a variable $y$ .
The \emph{select} statement (loosely corresponding to library functions such as \verb+iloc+, \verb+loc+, etc., in Python \textrm{pandas}) returns a subset data frame $y$ of $x$, based on an array of row indexes $\bar{r}$ and a set of column labels $C$ . 
The selection parameters $\bar{r}$ and $C$ are optional: when missing the selection includes all rows or columns of the given data frame.
The \emph{merge} statements are binary merge operations between data frames (the \emph{concat} and \emph{join} operations roughly match the default Python \textrm{pandas} \verb+concat+ and \verb+merge+ library functions, respectively).
%
%
The \emph{function} statements modify a data frame $x$ either by normalizing it (with the \emph{normalize} function) or by applying some \emph{other} function. The \emph{normalize} function produces a \emph{tainted} data frame $y$
(representing normalization functions such as standardization or scaling in Python \textrm{Sklearn}). We assume that any \emph{other} function does not produce tainted data frames.
Finally, \emph{use} statements employs data frames for either training ($f = \mbox{train}$) or testing ($f = \mbox{test}$) a ML model.

\subsubsection{Constructive Data Leakage Semantics}

We can now instantiate the definition of our data leakage semantics $\dot{\Lambda}_I$ with our small data frame-manipulating language. Given a program $P \equiv S_1, \dots, S_n$ written in our small language (where $S_1, \dots, S_n$ are statements), the set of input data frame variables $\inputs_P$ is given (with a slight abuse of notation, for simplicity) by the set of data files read by \emph{source} statements, i.e., 
$\inputs_P \defined i\traceS{P} = i\traceS{S_n} \circ \dots \circ i\traceS{S_1}\emptyset $, where $i\traceS{y = \text{read}(name)}I \defined I \cup \set{name}$ and $i\traceS{S}I \defined I$ for any other statement $S$ in $P$. 
Similarly, we define the set of used variables $\outputs_P \defined u\traceS{P} = u\traceS{S_n} \circ \dots \circ u\traceS{S_1}\emptyset$, where $u\traceS{f(X)}U \defined U \cup X$ and $u\traceS{S}U \defined U$ for any other statement $S$,
and analogously for $\otrain_P \subseteq \outputs_P$ (when $f = \mathrm{train}$) and $\otest_P \subseteq \outputs_P$ (when $f = \mathrm{test}$).

Our constructive data leakage semantics is $\dot{\collectingS{P}} \defined s\traceS{S_n} \circ \dots \circ s\traceS{S_1}\dot{\emptyset}$ where $\dot{\emptyset}$ is the totally undefined function and the semantic function $s\traceS{S}$ for each statement $S$ in $P$ is defined as follows:
\begin{align*}\label{eq:constructive}
        s\traceS{y = \text{read}(name)}m &\defined m\left[y \mapsto \lambda r \in R_{\text{read}()}\colon \set{name[r]}\right] \\
        s\traceS{y = x.\text{select}[\bar{r}][C]}m &\defined m\left[y \mapsto \lambda r \in R_{x.\text{select}[\bar{r}][C]}\colon m(x)(\bar{r}[r])\right] \\ 
        s\traceS{y = \text{concat}(x_1, x_2)}m &\defined \\
        & \hspace{-8em} m\left[ y \mapsto \lambda r \in R_{\text{concat}(x_1, x_2)}\colon {\begin{cases}
        m(x_1)r & r \leq |{\dom(m(x_1))}| \\
        m(x_2)(r\!-\!|{\dom(m(x_1))}|) & r > |{\dom(m(x_1))}|
        \end{cases}} \right] \\
        s\traceS{y = \text{join}(x_1, x_2)}m &\defined m\left[y \mapsto \lambda r \in R_{\text{join}(x_1, x_2)}\colon m(x_1)\overleftarrow{r} \cup m(x_2)\overrightarrow{r}\right] \\
        s\traceS{y = \text{normalize}(x)}m &\defined m\left[y \mapsto \lambda r \in R_{\text{normalize}(x)}\colon \bigcup_{r' \in \dom(m(x))} m(x)r'\right] \\
        s\traceS{y = \text{other}(x)}m &\defined m\left[y \mapsto \lambda r \in R_{\text{other}(x)}\colon m(x)r\right] \\
        s\traceS{\text{use}(x)}m &\defined m
\end{align*}
The semantics of the \emph{source} statement maps each row $r$ of a read data frame $y$ to (the set containing) the corresponding row in the read data file ($name[r]$).
The semantics of the \emph{select} statement maps each row $r$ of the resulting data frame $y$ to the set of data sources ($m(x)$) of the corresponding row ($\bar{r}[r]$) in the original data frame.
The \emph{concat} operation between two data frames $x_1$ and $x_2$ yields a data frame with all rows of $x_1$ followed by all rows of $x_2$. Thus, the semantics of \emph{concat} statements accordingly maps each row $r$ of the resulting data frame $y$ to the set of data sources of the corresponding row in $x_1$ (if $r \leq |\dom(m(x_1))|$, that is, $r$ falls within the size of $x_1$) or $x_2$ (if $r > |\dom(m(x_1))|$).
Instead, the \emph{join} operation combines two data frames $x_1$ and $x_2$ based on a(n index) column and yields a data frame containing only the rows that have a matching value in both $x_1$ and $x_2$. Thus, the semantics of \emph{join} statements maps each row $r$ of the resulting data frame $y$ to the union of the sets of data sources of the corresponding rows ($\overleftarrow{r}$ and $\overrightarrow{r}$) in $x_1$ and $x_2$.
We consider only one type of join operation (inner join) for simplicity, but other types (outer, left, or right join) can be similarly defined.
The normalize function is a tainting function so the semantics for the \emph{normalize} function introduces dependencies for each row $r$ in the normalized data frame $y$ with the data sources ($m(x)$) of each row $r'$ of the data frame before normalization. Instead, the semantics of \emph{other} (non-tainting) functions maintains the same dependencies ($m(x)r$) for each row $r$ of the modified data frame $y$.
Finally, \emph{use} statements do not modify any dependency so the semantics of \emph{use} statements leaves the dependencies map unchanged.

%% file: technical.tex
\section{Data Leakage Analysis}
\label{sec:analysis}

In this section, we abstract our concrete data leakage semantics to obtain a sound data leakage static analysis. In essence, our analysis keeps track of (an over-approximation of) the data source cells each data frame variable depends on (to detect potential explicit data source overlaps). Plus, it tracks whether data source cells are tainted, i.e., modified by a library function in such a way that could introduce data leakage (by implicit indirect data source overlaps).


\subsection{Data Sources Abstract Domain}

\subsubsection{Data Frame Abstract Domain.}\label{sec:ddom}

We over-approximate data sources by means of a parametric data frame abstract domain $\ddom(\cdom, \rdom)$, where the parameter abstract domains $\cdom$ and $\rdom$ track data sources columns and rows, respectively. We illustrate below two simple instances of these domains. 

\paragraph{Column Abstraction.}

We propose an instance of $\cdom$ that over-approximates the set of column labels in a data frame.
As, in practice, data frame labels are pretty much always strings, the elements of $\cdom$ belong to a complete lattice $\langle \celm, \sqsubseteq_C, \sqcup_C, \sqcap_C, \bot_C, \top_C \rangle$ where $\celm \defined \powerset{\mathbb{W}} \cup \set{\top_C}$; $\mathbb{W}$ is the set of all possible strings of characters in a chosen alphabet and $\top_C$ represents a lack of information on which columns a data frame \emph{may} have (abstracting any data frame). Elements in $\celm$ are ordered by set inclusion extended with $\top_C$ being the largest element: $C_1 \sqsubseteq_C C_2 \stackrel{\mbox{\tiny def}}{\Leftrightarrow} C_2 = \top_C \lor (C_1 \not= \top_C \land C_1 \subseteq C_2)$. Similarly, join $\sqcup_C$ and meet $\sqcap_C$ are set inclusion and set intersection, respectively, extended to account for $\top_C$:
\begin{equation*}
    \begin{aligned}
    C_1 \! \sqcup_C \! C_2 \defined \begin{cases}
    \top_C & C_1 \! = \! \top_C \lor C_2 \! = \! \top_2 \\
    C_1 \! \cup \! C_2 & \mbox{otherwise}
    \end{cases} &&
    C_1 \! \sqcap_C \! C_2 \defined \begin{cases}
    C_1 & C_2 = \top_C \\
    C_2 & C_1 = \top_C \\
    C_1 \! \cap \! C_2 & \mbox{otherwise}
    \end{cases}
    \end{aligned}
\end{equation*}
Finally, the bottom 
$\bot_C$ is 
the empty set $\emptyset$ (abstracting an empty data frame).


\paragraph{Row Abstraction.}

Unlike columns, data frame rows are not named. Moreover, data frames typically have a large number of rows and often ranges or rows are added to or removed from data frames. Thus, the abstract domain of intervals \cite{CousotC76} \emph{over the natural numbers} is a suitable instance of $\rdom$. The elements of $\rdom$ belong to the complete lattice $\langle \relm, \sqsubseteq_R, \sqcup_R, \sqcap_R, \bot_R, \top_R \rangle$ with the set $\relm$ defined as $\relm \defined \set{[l, u] \mid l \in \nat, u \in \nat \cup \set{\infty}, l \leq u} \cup \set{\bot_R}$. The top element $\top_R$ is $[0, \infty]$. Intervals in $\rdom$ abstract (sets of) row indexes: the concretization function $\gamma_R\colon \relm \rightarrow \powerset{\nat}$ is such that $\gamma_R(\bot_R) \defined \emptyset$ and $\gamma_R([l, u]) \defined \set{r \in \nat \mid l \leq r \leq u}$. The interval domain partial order ($\sqsubseteq_R$) and operators for join ($\sqcup_R$) and meet ($\sqcap_R$) are defined as usual (e.g., see Miné's PhD thesis \cite{Mine04} for reference).

In addition, we associate with each interval $R \in \relm$ another interval $\idx{R}$ of indices: $\idx{\bot_R} \defined \bot_R$ and $\idx{[l, u]} \defined [0, u - l]$; this essentially establishes a map $\phi_R\colon \nat \rightarrow \nat$ between elements of $\gamma_R(R)$ (ordered by $\leq$) and elements of $\gamma_R(\idx{R})$ (also ordered by $\leq$). In the following, given an interval $R \in \relm$ and an interval of indices $[i, j] \in \relm$ (such that $[i, j] \sqsubseteq_R R$), we slightly abuse notation and write $\phi^{-1}_R([i, j])$ for the sub-interval of $R$ between the indices $i$ and $j$, i.e., we have that $\gamma_R(\phi^{-1}_R([i, j])) \defined \set{r \in \gamma(R) \mid \phi^{-1}(i) \leq r \leq \phi^{-1}(j)}$. We need this operation to soundly abstract consecutive row selections 
(cf. Section~\ref{sec:analysis}). 
%


\begin{example}[Row Abstraction]
Let us consider the interval $[10, 14] \in \relm$ with index $\idx{R} = [0, 4]$. We have an isomorphism $\phi_R$ between $\set{10, 11, 12, 13, 14}$ and $\set{0, 1, 2, 3, 4}$. Let us consider now the interval of indices $[1, 3]$. We then have $\phi^{-1}_R([1, 3]) = [11, 13]$ (since $\phi^{-1}_R(1) = 11$ and $\phi^{-1}_R(3) = 13$).
\end{example}


\paragraph{Data Frame Abstraction.}

The elements of the data frame abstract domain $\ddom(\cdom, \rdom)$ belong to a partial order $\langle \delm, \sqsubseteq_L \rangle$ where $\delm \defined \mathbb{W} \times \celm \times \relm$ contains triples of a data file name $file \in \mathbb{W}$, a column over-approximation $C \in \celm$, and a row over-approximation $R \in \relm$. In the following, we write $\text{\footnotesize file}^C_R$ for the abstract data frame $\langle \text{\footnotesize file}, C, R \rangle \in \delm$.
The partial order $\sqsubseteq_L$ compares abstract data frames derived from the same data files: $X^C_R \sqsubseteq_L Y^{C'}_{R'} \stackrel{\mbox{\tiny def}}{\Leftrightarrow} X = Y \land C \sqsubseteq_C C' \land R \sqsubseteq R'$.

We also define a predicate 
for whether 
abstract data frames overlap: 
\begin{equation}\label{eq:overlap}
\mathrm{overlap}(X^C_R, Y^{C'}_{R'}) \stackrel{\mbox{\tiny def}}{\Leftrightarrow} X = Y \land C \sqcap_C C' \not= \emptyset \land R \sqcap R' \not= \bot_R
\end{equation}
and partial join ($\sqcup_L$) and meet ($\sqcap_L$) 
over data frames from the same data files: 
\begin{equation*}
    \begin{aligned}
    X^{C_1}_{R_1} \sqcup_L X^{C_2}_{R_2} \defined X^{C_1 \sqcup_C C_2}_{R_1 \sqcup_R R_2} &\qquad&
    X^{C_1}_{R_1} \sqcap_L X^{C_2}_{R_2} \defined X^{C_1 \sqcap_C C_2}_{R_1 \sqcap_R R_2}
    \end{aligned}
\end{equation*}

Finally, we define a constraining operator $\downarrow^C_R$ that restricts an abstract data frame to given column and row over-approximations: $X^C_R\downarrow^{C'}_{R'} \defined X^{C \sqcap_C C'}_{\phi^{-1}(\idx{R} \sqcap_R R')} $. Note that here the definition makes use of $\idx{R}$ and $\phi^{-1}([i, j])$ to compute the correct row over-approximation.

\begin{example}[Abstract Data Frames]
Let $\text{\footnotesize file}^{\set{id, city}}_{[10,14]}$ abstract a data frame with columns $\{id, city\}$ and rows $\{10, 11, 12, 13, 14\}$ derived from a data source $file$. The abstract data frame $\text{\footnotesize file}^{\set{country}}_{[12,15]}$ does not overlap with it, while $\text{\footnotesize file}^{\set{id}}_{[12,15]}$ does. Joining $\text{\footnotesize file}^{\set{id, city}}_{[10,14]}$ with $\text{\footnotesize file}^{\set{country}}_{[12,15]}$ yields 
$\text{\footnotesize file}^{\set{id, city, country}}_{[10,15]}$. Instead, the meet with 
$file^{\set{id}}_{[12,15]}$ 
yields 
$\text{\footnotesize file}^{\set{id}}_{[12,14]}$.
Finally, the constraining 
$\text{\footnotesize file}^{\set{id, city, country}}_{[10,15]}\downarrow^{\set{city}}_{[1,2]}$ results in $\text{\footnotesize file}^{\set{city}}_{[11,12]}$ (since $\phi^{-1}_{[10,15]}(1) = 11$ and $\phi^{-1}_{[10,15]}(2) = 12$).
\end{example}

In the rest of this section, for brevity, we simply write $\ddom$ instead of $\ddom(\cdom, \rdom)$.

\subsubsection{Data Frame Set Abstract Domain.}\label{sec:sdom}

Data frame variables may depend on multiple data sources. We thus lift our abstract domain $\ddom$ to an abstract domain $\sdom(\ddom)$ of sets of abstract data frames.
The elements of $\sdom(\ddom)$ belong to a lattice $\langle \selm, \sqsubseteq_S, \sqcup_S, \sqcap_S \rangle$ with $\selm \defined \powerset{\delm}$. Sets of abstract data frames in $\selm$ are maintained in a \emph{canonical form} such that no abstract data frames in a set can be overlapping (cf. Equation~\ref{eq:overlap}). The partial order $\sqsubseteq_S$ between canonical sets relies on the partial order between abstract data frames: $S_1 \sqsubseteq_S S_2 \stackrel{\mbox{\tiny def}}{\Leftrightarrow} \forall L_1 \in S_1 \exists L_2 \in S_2\colon L_1 \sqsubseteq_L L_2$.

The join ($\sqcup_S$) and meet ($\sqcap_S$) operators perform a set union and set intersection, respectively, followed by a reduction 
to make the result canonical:
\begin{equation*}
    \begin{aligned}
    S_1 \sqcup_S S_2 \defined \mbox{\textsc{reduce}}^{\sqcup_L}(S_1 \cup S_2) &\qquad&
    S_1 \sqcap_S S_2 \defined \mbox{\textsc{reduce}}^{\sqcap_L}(S_1 \cap S_2)
    \end{aligned}
\end{equation*}
where $\mbox{\textsc{reduce}}^{op}(S) \defined \set{L_1 op L_2 \mid L_1, L_2 \in S, \mathrm{overlap}(L_1, L_2)} ~\cup \\ \set{L_1 \in S \mid \forall L_2 \in S \setminus \set{L_1}\colon \neg\mathrm{overlap}(L_1, L_2)}$

Finally, we lift 
$\downarrow^{C}_{R}$ by element-wise application: $S\downarrow^{C}_{R} \defined \set{L\downarrow^{C}_{R} \mid L \in S}$. 

\begin{example}[Abstract Data Frame Sets]
Let us consider the join of two abstract data frame sets $S_1 = \left\{ \text{\footnotesize file1}_{[1, 10]}^{\{id\}}, \text{\footnotesize file2}_{[0, 100]}^{\{ name \}} \right\}$ and 
$S_2 = \left\{ \text{\footnotesize file1}_{[9, 12]}^{\{ id\}}, \text{\footnotesize file3}_{[0, 100]}^{\{ zip \}} \right\}$ 
Before reduction, we obtain 
$\left\{ \text{\footnotesize file1}_{[1, 10]}^{\{id\}}, \text{\footnotesize file1}_{[9, 12]}^{\{id\}}, \text{\footnotesize file2}_{[0, 100]}^{\{ name \}} , \text{\footnotesize file3}_{[0, 100]}^{\{ zip \}} \right\}$.
The reduction operation makes the set canonical:
$\left\{ \text{\footnotesize file1}_{[1, 12]}^{\{id\}}, \text{\footnotesize file2}_{[0, 100]}^{\{ name \}} , \text{\footnotesize file3}_{[0, 100]}^{\{ zip \}} \right\}$.
\end{example}

In the following, for brevity, we omit $\ddom$ and simply write $\sdom$ instead of $\sdom(\ddom)$.

\subsubsection{Data Frame Sources Abstract Domain.}\label{sec:adom}

We can now define the domain $\vars \rightarrow \adom(\sdom)$ that we use for our data leakage analysis. Elements in this abstract domain are maps from data frame variables in $\vars$ to elements of a data frame sources abstract domain $\adom(\sdom)$, which over-approximates the (input) data frame variables (indexed at some row) from which a data frame variable depends on.

In particular, elements in $\adom(\sdom)$ belong to a lattice $\langle \aelm, \sqsubseteq_A, \sqcup_A, \sqcap_A, \bot_A \rangle$ where $\aelm \defined \selm \times \mathbb{B}$ contains pairs $\langle S, B \rangle$ of a data frame set abstraction in $S \in \selm$ and a boolean flag in $B \in \mathbb{B} \defined \set{\mbox{\textsc{untaninted}}, \mbox{\textsc{maybe-tainted}}}$.
In the following, given an abstract element $m \in \vars \rightarrow \aelm$ of $\vars \rightarrow \adom(\sdom)$ and a data frame variable $x \in \vars$, we write $m_s(x) \in \selm$ and $m_b(x) \in \mathbb{B}$ for the first and second component of the pair $m(x) \in \aelm$, respectively.

The abstract domain operators apply component operators pairwise: $\sqsubseteq_A \defined \sqsubseteq_S \times \leq$, $\sqcup_A \defined \sqcup_S \times \lor$, $\sqcap_A \defined \sqcap_S \times \land$, 
where 
$\leq$ in $\mathbb{B}$ is such that $\mbox{\textsc{untainted}} \leq \mbox{\textsc{maybe-tainted}}$. The bottom element $\bot_A$ is $\langle \emptyset, \textsc{untainted} \rangle$.

Finally, we define the concretization 
$\gamma\colon (\vars \rightarrow (\nat \rightarrow \aelm)) \rightarrow (\vars \rightarrow (\nat \rightarrow \powerset{\vars \times \nat}))$:
    $\gamma(m) \defined \lambda x \in \vars\colon \left( \lambda r \in \nat\colon \gamma_A(m(x)) \right)$, 
where $\gamma_A\colon \aelm \rightarrow \powerset{\vars \times \nat}$ is 
    $\gamma_A(\langle S, B \rangle) \defined \set{ X[r] \mid X^C_R \in S, r \in \gamma_R(R) }$   
(with $\gamma_R\colon \relm \rightarrow \powerset{\nat}$ being the concretization function for row abstractions, cf. Section~\ref{sec:ddom}). Note that, $\gamma_A$ does not use $B \in \mathbb{B}$ nor $C \in \celm$. These are devices uniquely needed by our abstract semantics (that we define below) to track (and approximate the concrete actual) dependencies across program statements.

\subsection{Abstract Data Leakage Semantics}\label{subsec:analysis}

Our data leakage analysis is given by $\dot{\collectingS{P}^\natural} \defined a\traceS{S_n} \circ \dots \circ a\traceS{S_1}\dot{\bot_A}$ where $\dot{\bot_A}$ maps all data frame variables to $\bot_A$ and the abstract semantic function $a\traceS{S}$ for each statement in $P$ is defined as follows:
\allowdisplaybreaks
    \begin{align*}
        a\traceS{y = \text{read}(name)}m &\defined m\left[y \mapsto \langle \set{name^{\top_C}_{[0, \infty]}}, \textsc{false} \rangle \right] \\
        a\traceS{y = x.\text{select}[\bar{r}][C]}m &\defined m\left[y \mapsto \begin{cases}
        \langle
        m_s(x)\downarrow^{C}_{[{\scriptstyle\mbox{min}}(\bar{r}), {\scriptstyle\mbox{max}}(\bar{r})]}, m_b(x)
        \rangle & \neg m_b(x) \\
        \langle m_s(x), m_b(x) \rangle & \mbox{otherwise}
        \end{cases}
        \right] \\ 
        a\traceS{y = \text{op}(x_1, x_2)}m &\defined m\left[ y \mapsto \langle
        m_s(x_1) \sqcup_S m_s(x_2), m_b(x_1) \lor m_b(x_2)
        \rangle\right] \\
        a\traceS{y = \text{normalize}(x)}m &\defined m\left[y \mapsto \langle
        m_s(x), \textsc{true}
        \rangle\right] \\
        a\traceS{y = \text{other}(x)}m &\defined m\left[y \mapsto \langle
        m_s(x), m_b(x)
        \rangle\right] \\
        a\traceS{\text{use}(x)}m &\defined m
    \end{align*}
The abstract semantics of the \emph{source} statement simply maps a read data frame variable $y$ to the untainted abstract data frame set containing the abstraction of the read data file ($name^{\top_C}_{[0, \infty]}$). 
The abstract semantics of the \emph{select} statement maps the resulting data frame variable $y$ to the abstract data frame set $m_s(x)$ associated with the original data frame variable $x$; in order to soundly propagate (abstract) dependencies, $m_s(x)$ is constrained by  $\downarrow^{C}_{[{\scriptstyle\mbox{min}}(\bar{r}), {\scriptstyle\mbox{max}}(\bar{r})]}$ (cf. Section~\ref{sec:sdom}) only if $m_s(x)$ is untainted.
The abstract semantics of \emph{merge} statements merges the abstract data frame sets $m_s(x_1)$ and $m_s(x_2)$ and taint flags $m_b(x_1)$ and $m_b(x_2)$ associated with the given data frame variables $x_1$ and $x_2$. 
Note that such semantics is a sound but rather imprecise abstraction, in particular, for the \emph{join} operation. More precise abstractions can be easily defined, at the cost of also abstracting data frame contents.
The abstract semantics of \emph{function} statements maps the resulting data frame variable $y$ to the abstract data frame set $m_s(x)$ associated with the original data frame variable $x$; the \emph{normalize} function sets the taint flag to \textsc{true}, while \emph{other} functions leave the taint flag $m_b(x)$ unchanged.
Note that, unlike the analysis sketched by Subotić et al.~\cite{NBLyzer}, we do not perform any renaming or resetting of the data source mapping (cf. Section 4.2 in \cite{NBLyzer}) but we keep tracking dependencies with respect to the input data frame variables.
Finally, the abstract semantics of \emph{use} statements leave the abstract dependencies map unchanged. 

The abstract data leakage semantics $\dot{\collectingS{P}^\natural}$ is \emph{sound}: 

\begin{theoremrep}
\label{thm:soundness}
$P \models \noleakage \Leftarrow \gamma(\dot{\collectingS{P}^\natural}) ~\dot{\supseteq}~ \dot{\alpha}(\alpha_{\dependency^+}(\noleakage))$
\end{theoremrep}
 \begin{appendixproof}[Sketch]
 The proof follows from the definition of abstract data leakage semantics $\dot{\collectingS{P}^\natural}$ and that of the concretization function $\gamma$, 
 observing that all abstract semantic functions $a\traceS{S}$ for a statement $S$ in $P$ always over-approximate the set of input data sources from which a data frame variable depends on. 
 \end{appendixproof}

Similarly, we have the sound but not complete counterpart of Lemma~\ref{lm:check} for practically checking absence of data leakage:
\begin{lemma}\label{lm:acheck}
\begin{equation*}
\begin{aligned}
P \models \noleakage &\Leftarrow \forall o_1 \in \otrain_P, o_2 \in \otest_P\colon \\
&\hspace{1.8em}
\bigcup_{r_1 \in \dom(\gamma(\dot{\collectingS{P}^\natural})o_1)} \gamma(\dot{\collectingS{P}^\natural})o_1(r_1) 
~\cap~ 
\!\!\!\!\!\!\!\!
\bigcup_{r_2 \in \dom(\gamma(\dot{\collectingS{P}^\natural})o_2)} \dot{\gamma(\collectingS{P}^\natural})o_2(r_2) = \emptyset \\
&\Leftrightarrow \forall o_1 \in \otrain_P, o_2 \in \otest_P\colon 
\forall X^{C}_{R} \in \dot{\collectingS{P}^\natural_s}o_1, Y^{C'}_{R'} \in \dot{\collectingS{P}^\natural_s}o_2\colon 
\\
&\hspace{1.8em} 
\neg \mathrm{overlap}(X^C_R, Y^{C'}_{R'}) 
\land \left(X = Y \Rightarrow \neg \dot{\collectingS{P}^\natural_b}o_1 \land \neg \dot{\collectingS{P}^\natural_b}o_2 \right) 
\end{aligned}
\end{equation*}
\end{lemma}
Specifically, Lemma~\ref{lm:acheck} allows us to verify  absence of data leakage by checking that any pair of abstract data frames sources $X^{C}_{R}$ and $Y^{C'}_{R'}$ for data respectively used for training (i.e., $o_1$) and testing (i.e., $o_2$) are disjoint (i.e, $\neg \mathrm{overlap}(X^C_R, Y^{C'}_{R'})$, cf. Equation~\ref{eq:overlap}) and untainted (i.e., $\neg \dot{\collectingS{P}^\natural_b}o_1 \land \neg \dot{\collectingS{P}^\natural_b}o_2$ if $X = Y$, that is, if they originate from the same data file).

\begin{example}[Motivating Example (Continued)]
The data leakage analysis of our motivating example (cf. Example~\ref{ex:notebook}) is the following:
\begin{flalign*}
\scriptstyle  
& \scriptstyle a\bigl\llbracket\footnotesize\texttt{data = read(\textcolor{mygreen}{"data.csv"})}\bigr\rrbracket\dot{\bot_A} = 
\left( m_1 \defined \lambda x\colon \begin{cases}
\scriptstyle \left\langle \set{\text{\footnotesize data.csv}^{\top_C}_{[0,\infty]}}, \mbox{\textsc{false}} \right\rangle & \displaystyle x = data \\
\scriptstyle \mbox{undefined} & \scriptstyle \mbox{otherwise}
\end{cases} \right) & \\
& \scriptstyle a\bigl\llbracket\footnotesize \texttt{X = data.select[][\{\textcolor{mygreen}{"X\_1"}, \textcolor{mygreen}{"X\_2"}, \textcolor{mygreen}{"y"}\}]}\bigr\rrbracket m_1 = \\
& \scriptstyle \qquad \left( m_2 \defined m_1\left[ X \mapsto \left\langle \set{\text{\footnotesize data.csv}^{\set{``X\_1", ``X\_2", ``y''}}_{[0,\infty]}}, \mbox{\textsc{false}} \right\rangle \right] \right) & \\
& \scriptstyle a\bigl\llbracket\footnotesize \texttt{X\_norm = normalize(X)}\bigr\rrbracket m_2 = 
\left( m_3 \defined m_2\left[ X\_\mathrm{norm} \mapsto \left\langle \set{\text{\footnotesize data.csv}^{\set{``X\_1", ``X\_2", ``y''}}_{[0,\infty]}}, \mbox{\textsc{true}} \right\rangle
\right]\right) \\
& \scriptstyle a\bigl\llbracket\footnotesize \texttt{X\_train = X\_norm.select[[$\lfloor 0.025 * R_{\texttt{X\_norm}} \rfloor$ + 1, $\dots$, $R_{\texttt{X\_norm}}$]][]}\bigr\rrbracket m_3 = \\
& \scriptstyle \qquad \left( m_4 \defined m_3\left[ X\_\mathrm{train} \mapsto \left\langle \set{{\text{\footnotesize data.csv}}_{[\lfloor 0.025 * R_{X\_\mathrm{norm}} \rfloor + 1, R_{X\_\mathrm{norm}}]}^{\set{``X\_1", ``X\_2", ``y''}}}, \mbox{\textsc{true}} \right\rangle \right]\right) \\
& \scriptstyle a\bigl\llbracket\footnotesize \texttt{X\_test = X\_norm.select[[0, $\dots$, $\lfloor 0.025 * R_{\texttt{X\_norm}} \rfloor$]][]}\bigr\rrbracket m_4 = \\
& \scriptstyle \qquad \left( m_5 \defined m_4\left[ X\_\mathrm{test} \mapsto \left\langle \set{{\text{\footnotesize data.csv}}_{[0, \lfloor 0.025 * R_{X\_\mathrm{norm}} \rfloor]}^{\set{``X\_1", ``X\_2", ``y''}}}, \mbox{\textsc{true}} \right\rangle \right]\right) \\
%
& \scriptstyle a\bigl\llbracket\footnotesize \texttt{test(X\_test)}\bigr\rrbracket (a\bigl\llbracket\footnotesize \texttt{train(X\_train)}\bigr\rrbracket m_5) = m_5
\end{flalign*}
At the end of the analysis, $X\_\mathrm{train} \in \otrain$ and $X\_\mathrm{test} \in \otest$ depend on disjoint but \emph{tainted} abstract data frames derived from the same input file $data.csv$. Thus, the absence of data leakage check from Lemma~\ref{lm:acheck} (rightfully) fails.
\end{example}

%% file: evaluation.tex
\section{Experimental Evaluation}
\label{sec:evaluation}

We implemented our static analysis into the open source 
\textsc{NBLyzer}~\cite{NBLyzer} framework for data science notebooks. 
\textsc{NBlyzer} 
performs the analysis starting on an individual code cell (\emph{intra-cell analysis}) and, based on the resulting abstract state, it proceeds to analyze \emph{valid} successor code cells (\emph{inter-cell analysis}). Whether a code cell is a valid successor or not, is specified by an analysis-dependent \emph{cell propagation condition}. We refer to the original \textsc{NBLyzer} paper~\cite{NBLyzer} and to Appendix~\ref{app:nblyzer} and~\ref{sec:integration} for further details.

\begin{wraptable}{R}{0.4\linewidth}
\footnotesize
\vspace{-1.5em}
\caption{Alarms raised by the previous \cite{NBLyzer} vs our analysis.}
\centering
\label{tab:pres}
\begin{tabular}{c||cc|c}
\hline
\multirow{2}{*}{Analysis}   & \multicolumn{2}{c|}{TP} & \multirow{2}{*}{FP} \\ 
          & Taint & Overlap &    \\ \hline \hline
\cite{NBLyzer}  & 10     & 0       & 2  \\ \hline
\textbf{Ours} & \textbf{10}     & \textbf{15}       & \textbf{2} \\ \hline
\end{tabular}
\vspace{-1.5em}
\end{wraptable}

We evaluated\footnote{Experiments done on a Ryzen 9 6900HS with 24GB DDR5 running Ubuntu 22.04.} our analysis against the data leakage analysis previously implemented in NBLyzer~\cite{NBLyzer}, using the same benchmarks. These are notebooks written to succeed in non-trivial real data science tasks and can be assumed to closely represent code written by non-novice data scientists.

%

We summarize the results in Table~\ref{tab:pres}. For each reported alarm, we engaged 4 data scientists at Microsoft to determine true (TP) and false positives (FP). We further classified the true positives as due to a normalization taint (Taint) or overlapping data frames (Overlap).
%
%
Our analysis found $10$ Taint data leakages in $5$ notebooks, and $15$ Overlap data leakage in $11$ notebooks, i.e., a $1.2\%$ bug rate, which adheres to true positive bug rates reported for null pointers in industrial settings (e.g., see \cite{inferstudy}). 
The previous analysis only found $10$ Taint  leakages in $5$ notebooks. It could not  Overlap data leakages because it cannot reason at the granularity of partial data frames.
The cost for our more precise analysis is a mere $7\%$ slowdown.
%
Both analyses reported $2$ false positives, 
due to different objects having the same function name (e.g., 
 \texttt{LabelEncoder} and \texttt{StandardScaler} both having the function \texttt{fit\_transform}). This issue can be solved by introducing object sensitive type tracking in our analysis. We leave it for future work.

The full experimental evaluation is described in Appendix~\ref{apx:evaluation}.

%% file: related.tex
\section{Related Work}
\label{sec:rel}

\paragraph{\bf Related Abstract Interpretation Frameworks and Static Analyses.}

As mentioned in Section~\ref{sec:semantics}, our framework generalizes the notion of data usage proposed by Urban and Müller~\cite{input} and the definition of dependency abstraction used by Cousot~\cite{Cousot19}. In particular, among other things, the generalization involves reasoning about dependencies (and thus data usage relationships) between multi-dimensional variables. In \cite{input}, Urban and Müller show that information flow analyses can be used for reasoning about data usage, albeit with a loss in precision unless one repeats the information flow analysis by setting each time a different input variable as high security variable (cf. Section 8 in \cite{input}). In virtue of the generalization mentioned above, the same consideration applies to our work, i.e., information flow analyses can be used to reason about data leakage, but with an even higher cost to avoid a precision loss (the analysis needs to be repeated each time for different portions of the input data sources). Analogously, in \cite{Cousot19}, Cousot shows that information flow, slicing, non-interference, dye, and taint analyses are all further abstractions of his proposed framework (cf. Section 7 in \cite{Cousot19}). As such, they are also further (and thus less precise) abstractions of our proposed framework. In particular, a taint analysis will only be able to detect a subset of the data leakage bugs that our analysis can find, i.e., those solely originating from library transformations but not those originating from (partially) overlapping data. Vice versa, our proposed analysis could also be used as a more fine-grained information flow or taint analysis. 

\paragraph{\bf Static Analysis for Data Science.}
Static analysis for data science is an emerging area in the program analysis 
community~\cite{Urban19}. Some notable static analyses for data science scripts include an analysis for ensuring correct shape dimensions in TensorFlow programs~\cite{yannis}, an analysis for constraining inputs based on program constraints~\cite{input}, and a provenance analysis~\cite{vamsa}. In addition, static analyses have been proposed for 
data science notebooks~\cite{NBLyzer,nbsafety}.  \textsc{NBLyzer}~\cite{NBLyzer} contains an ad-hoc data leakage analysis that detects Taint data leakages~\cite{NBLyzer}. 
An extension to handle Overlap data leakages was sketched in~\cite{SuboticBS22} (but no analysis precision results were reported). However, both these analysis are not formalized nor formally proven sound\footnote{We found a number of soundness issues in~\cite{SuboticBS22} when working on our formalization.}. In contrast,  we introduce a sound data leakage analysis that has a rigorous semantic underpinning.

%% file: conclusion.tex
\section{Conclusion}
\label{sec:conclusion}
We have presented an approach for detecting data leakages statically. 
We provide a formal and rigorous derivation from the standard program collecting semantics, via successive abstraction to a final sound and computable static analysis definition. We  implement our analysis in the \textsc{NBLyzer} framework and demonstrate clear improvements upon previous ad-hoc data leakage analyses.

%% file: appendix.tex
\section{NBLyzer Framework}\label{app:nblyzer}
\looseness=-1
\textsc{NBlyzer} is designed specifically to adapt to the unique out-of-order execution semantics of notebooks. It performs the analysis starting on an individual code cell (\emph{intra-cell analysis}) and, based on the resulting abstract state, it proceeds to analyze \emph{valid} successor code cells (\emph{inter-cell analysis}). Whether a code cell is a valid successor or not, is specified by an analysis-dependent \emph{cell propagation $\phi$-condition}.

\begin{figure}[t]
    \centering
    \includegraphics[width=0.85\textwidth]{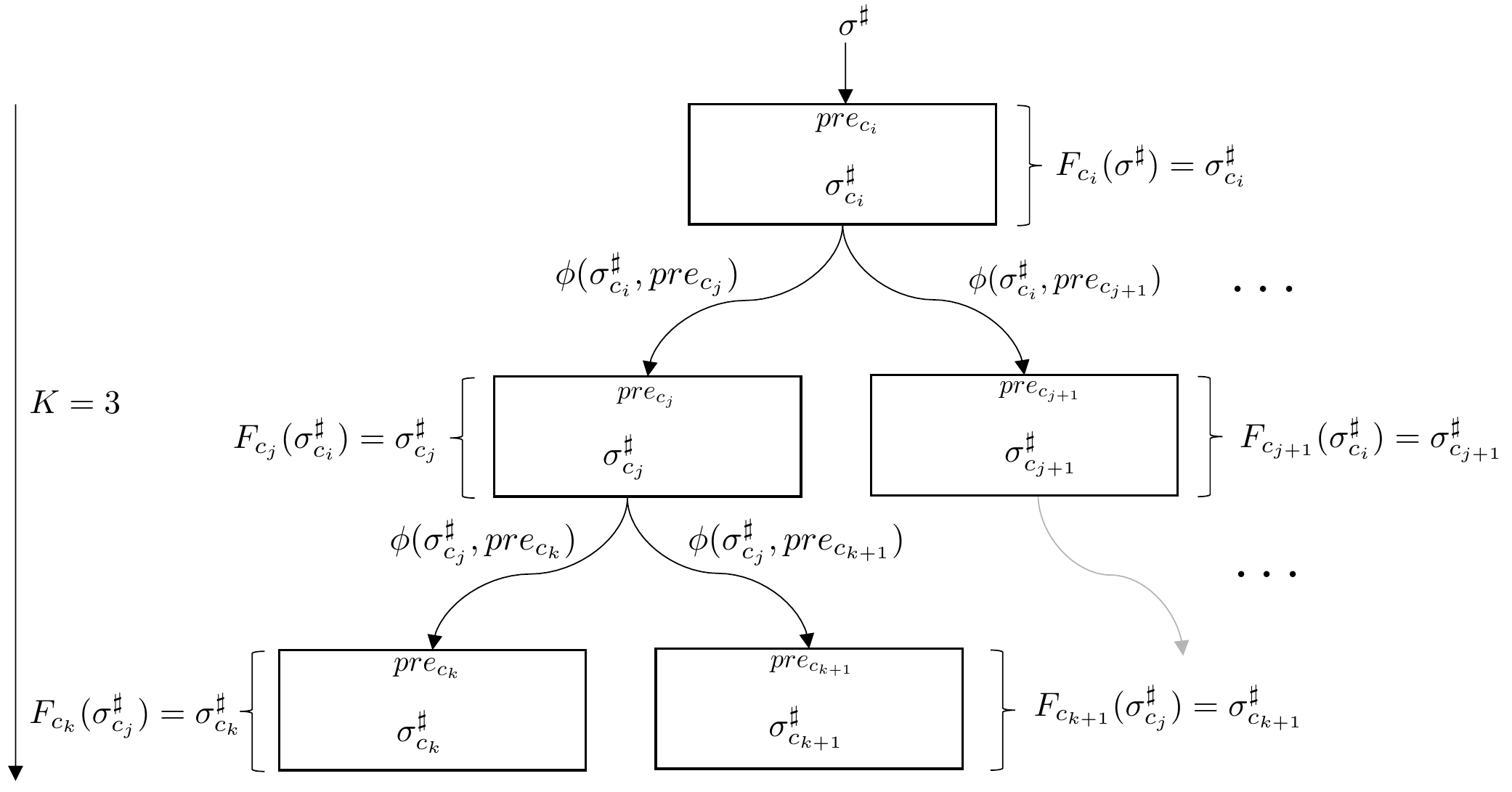}
    \caption{Inter-cell analysis\label{fig:fw}}
\end{figure}

Let $F_{c}$ be the abstract transformer that performs the analysis of an individual code cell $c$. The inter-cell analysis process of \textsc{NBlyzer} is visualized by the propagation tree in Figure~\ref{fig:fw}. At the intra-analysis level of each code cell $c$, the abstract transformer $F_{c}$ is applied to the current abstract state $\sigma^\sharp$ and returns an updated abstract state, i.e., $F_{c}(\sigma^\sharp) = \sigma^{\sharp\prime}$. The updated abstract state is propagated from one cell $c$ to another cell $c'$ if the $\phi$-condition holds. The $\phi$-condition depends on the incoming abstract state $F_{c}(\sigma^\sharp) = \sigma^{\sharp\prime}$ and a \emph{cell pre-condition} $pre_{c'}$ for $c'$. The cell pre-condition contains, e.g., unbound variables (i.e., variables used but not defined within the cell) and namespaces for libraries. The cell pre-conditions used in our analysis are specified in Section~\ref{sec:integration}.

Each propagation branch may terminate due to the following four cases: 
\begin{enumerate}
    \item the depth (number of individual code cells) of the propagation reaches a given (finite) \emph{propagation bound} $K \in \nat$;
    \item the $\phi$-condition does not hold for all code cells in the notebook;
    \item a fixpoint subsumption occurred: the $n^{th}$ time, $n > 1$, a cell was analyzed does not result in a change in the abstract state;
    \item a error e.g., a data leakage, has been detected that halts the propagation.
\end{enumerate}
It is also possible to not specify a finite propagation bound, i.e., $K = \infty$; in this case, condition (1) is ignored.
When all propagation branches terminate, the inter-cell analysis terminates.

\section{Implementation} 
\label{sec:integration}
We integrate our static analysis based on our approach described in Section~\ref{sec:analysis} into the open source 
\textsc{NBLyzer}~\cite{NBLyzer} framework for data science notebooks. 

\textsc{NBLyzer} provides the fixpoint machinery to execute an abstract semantics on a notebook as well as a reference data leakage analysis that we aim to improve on in this paper. Please refer to Appendix~\ref{app:nblyzer} for an overview of \textsc{NBLyzer}. Below we describe the main tasks required to integrate our data leakage analyzer into \textsc{NBLyzer} as well as several limitations of our implementation. 

\paragraph{Cell Propagation ($\phi$):}
In the \textsc{NBLyzer} framework each analysis, aside from implementing an abstract domain and abstract transfer functions, needs to define a $\phi$-condition to determine if propagation should continue to another cell. To achieve good performance, $\phi$ must be defined as strong as possible while not 
sacrificing soundness i.e., we do not want to miss any interesting execution sequences (e.g., containing a bug) by terminating prematurely. 

Moreover, each cell has a set of pre-condition variables $pre$ which we define as a subset of of unbound variables and namespaces that are used to invoke functions in the knowledge base or propagated to other cells. This includes include namespaces for libraries. For our data leakage analysis, only namespaces that relate to functions in our knowledge base (see below) are considered. We therefore specify the $\phi$-condition for inter-cell propagation as follows:
\begin{equation*}
\phi(m, pre_{c}) \defined pre_{c} \subseteq  \set{v \in {\mathrm{dom}}({m}) \mid X_{R}^{C} \in m(v), R \neq \bot } \land pre_{c} \neq \emptyset
\end{equation*}
where $m = \sigma^\sharp_{c}$ is the abstract state resulting from the analysis of the individual code cell $c$.
This rule stipulates the condition by which a successor cell should be analyzed. That is, if any variable that has rows (not $\bot$) in 
the abstract state of the current notebook cell, is also unbound in the successor notebook cell, we proceed to propagate the abstract state.


\paragraph{Support for Functions:} We support inter-procedural analysis via function inlining/cloning. If a function has been in an executed cell, we inline its body in any subsequent call site before processing the cell. In the case the definition does not exist in a predecessor cell, we treat the function as a undefined function.

\section{Extended Evaluation}\label{apx:evaluation}
\subsection{Experimental Setup}
\subsubsection{Environment}
All experiments were performed on a Ryzen 9 6900HS with 24GB DDR5 running Ubuntu 22.04. Python 3.10.6 was used to execute both versions of \textsc{NBLyzer}, one running our data leakage static analysis and the other the data leakage analysis from~\cite{NBLyzer}. We used default \textsc{NBLyzer} settings (e.g., $K = 5$).

\subsubsection{Benchmarks}
We use a data science notebook benchmark suite consisting of 4 Kaggle competitions
 that has previously been used to evaluate data science static 
analyzers~\cite{vamsa,NBLyzer}. We analyzed $2111$ over $2413$ notebooks, resulting in $7378$ notebook executions.  We excluded $302$ notebooks because they could not be digested by our analyzer (i.e., syntax errors, JSON decoding errors, etc.).  The benchmark characteristics are summarized in Table~\ref{tbl:bm}. 

All notebooks are written to succeed in a non-trivial data science competition task and can be assumed to closely represent code of non-novice data scientists. On average the notebooks in the benchmark suite have $24$ cells, where each cell on average has 
$9$ lines of code. On average, branching instructions appear in $33$\% of cells. Each notebook has on 
average $3$ functions and $0.1$ classes defined.

\begin{table}[t]
\caption{Kaggle Notebook Benchmark Characteristics\label{tbl:bm}}
\footnotesize
\centering 
\begin{tabular}{|l|l|l|l|l|}
Characteristic                    & Mean  & SD  & Max  & Min \\ \hline
Cells (per-notebook)              & 23.58 & 20.21  & 182  & 1 \\
Lines of code (per-cell)          & 9.12 & 13.55  & 257  & 1 \\
Branching inst. (per-cell) & 0.43  & 2.49 & 76 & 0 \\
Functions (per-notebook)          & 3.33 & 7.11 & 72 & 0 \\
Classes (per-notebook)            & 0.14  &0.64  & 11 & 0 \\
parse error cell (per-notebook)  & 0.5 & 0.98 & 20 & 0 \\
Variables (per-cell)              & 8.2 & 2.3 & 552 & 0 \\
Unbound variables (per-cell)      & 2.1 & 1.06 & 12 & 0 \\ \hline
\end{tabular}
\end{table}

\subsubsection{Use Case} 
\label{ssec:usecase}
As described in~\cite{NBLyzer}, the use case is a data leakage detector for notebooks in an Integrated Development Environment (IDE). The analysis is triggered by an event (such as a cell execution) and provides 
a \emph{what-if} analysis, namely: \emph{"if you do this event then the following future cell executions may lead to a data leakage"}. The analyzer should identify data leakages with a \emph{soft} analysis deadline of $1$ second in accordance to the RAIL performance model\footnote{\url{https://web.dev/rail/}}. 

\subsubsection{Experimental methodology}
For every notebook we perform our analysis for a 
\emph{valid execution}. We define a valid execution 
as a non-empty sequence of cells that starts with a cell that does not 
contain unbound variables and thus can commence a valid execution.

\subsection{Extended Precision Evaluation}

To further provide the reader with an appreciation of how data leakages manifest in our benchmarks, a code snippet showing an overlapping data error (in contrast to the normalization taint error in our motivating example) discovered from our benchmarks is shown in Figure~\ref{ex:victim}. Here repeated training and testing occurs on the same training set. In the code, a data frame is first loaded to a variable called \texttt{df}. After several exploratory data analysis (EDA) steps (not shown), the features (columns which are inputs to the model) and the target (prediction column) is split into two data frames, called \texttt{X} and \texttt{y} respectively. These two data frame variables are split into four (\texttt{X\_train}, \texttt{X\_test}, \texttt{y\_train}, \texttt{y\_test}) reflecting the training and testing subsets. The location of the split is defined by the \texttt{split} variable. The programmer, appears unsure about the semantics of \texttt{iloc} and assumes the end point needs to be incremented to 
obtain the right split index. As a consequence the data is split such that both training and testing contain the row at the value of \texttt{split}. The model, represented by the \texttt{lr\_clf} variable is created and the training is performed (invocation of \texttt{fit} method), using the train suffixed data frames. In the last cell, prediction is made (invocation of \texttt{predict} function) however with overlapping data. We note, this type of bug cannot be detected using a taint analysis. We elaborate on this further in Section~\ref{sec:rel}.

\setcounter{ipythcntr}{4}

\begin{figure}
\centering
\textit{\footnotesize Imports, initialization etc.} \\
...
\begin{pyin2}[cell1]
df = pd.read_csv("heart.csv")
\end{pyin2}

...\\

\setcounter{ipythcntr}{7}
\begin{pyin2}[cell3]
y = df[['target']]
X = df.drop('target', axis=1)

X_train = X.iloc[:split+1] 
X_test = X.iloc[split:end]

y_train = y.iloc[:split+1]
y_test = y.iloc[split:end]
 
\end{pyin2}

\begin{pyin2}[cell4]
lr_clf = LogisticRegression(solver='liblinear')
train1 = lr_clf.fit(X_train, y_train)
\end{pyin2}

\begin{pyin2}[cell5]
train_score = accuracy_score(y_test, lr_clf.predict(X_test))
\end{pyin2}
\caption{Example of overlapping data bug}
\label{ex:victim}
\end{figure}

\begin{wrapfigure}{R}{0.5\linewidth}
    \centering
    \begin{tikzpicture}
        \begin{axis}[
            area style,
            xtick={1,2, 3, 4, 5},
            xlabel={Error Path Length (No. cells)},
            ylabel={Frequency},
            width=0.35\textwidth,
            height=0.2\textwidth,
            ]
        \addplot+[ybar, mark=no] plot coordinates { 
            (1,6)
            (2,2)
            (4,5)
            (5,12)
        };
        \end{axis}
    \end{tikzpicture}
    \caption{Error path length frequency}
    \label{fig:errors}
\end{wrapfigure}
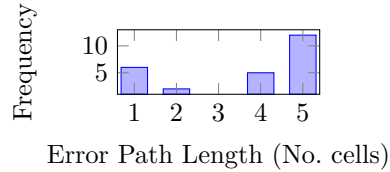

While our formalization in Section~\ref{sec:semantics} is sound w.r.t. our simple example language. We report several sources of unsoundness in our actual analysis implementation. Firstly, we do not support every 
Python 3 construct and instead focus on the most common constructs observed in notebook code. Constructs such as dynamic code evaluation, reference aliasing are not supported. We remark that data science Python programs are relatively simple compared to general Python programs. They tend to have largely linear control-flow and generally have a single call-site per function making cloning/inlining a reasonable choice. From our benchmarks we could only detect that $0.5\%$ of notebooks required an alias analysis (e.g., assigning data frames by reference) and for this reason we did not integrate an alias analysis. We find that it is infeasible to manually inspect all notebooks, and thus it is impractical to produce a recall rate for our benchmarks. 

The data leakages found had cell execution traces varied between $1-5$ cells in length indicating that the majority of bugs (76\%) manifested over several notebook cells, in a non-sequential cell order. We summarise these findings in the histogram in Figure~\ref{fig:errors}.

Given a simple knowledge base the coverage of our analysis is subject to an adequate formulation of the knowledge base which is elementary in our implementation, following the implementation in~\cite{NBLyzer}. Moreover, we believe lower-quality notebooks e.g., university subject assignments, may contain more data leakages. We leave additional modeling and precision investigation on other benchmarks for future work. 

We also note that we were not able to find instances that required widening and did not experience infinite (or very high) ascending chains being computed when analyzing our benchmarks. For this to occur we would need code that iterates on data frame rows, which is not common. Standard interval widening~\cite{CC77} can be applied if required.

\subsection{Performance Evaluation}
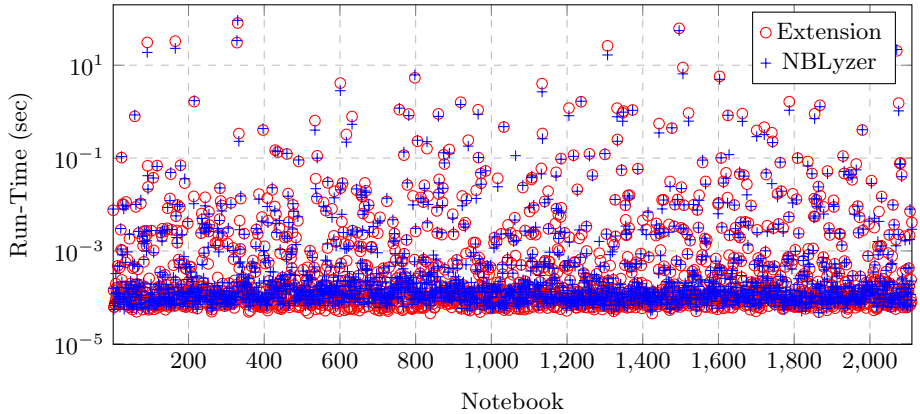
\begin{figure*}
\begin{tikzpicture}
\begin{axis}[
width=\textwidth,
height=0.5\textwidth,
xlabel={Notebook},
ylabel={Run-Time (sec)},
ymode=log,
xmin=1, xmax=2111, 
ymin=0, ymax=200,
xmajorgrids=true,
ymajorgrids=true,
grid style=dashed,
scaled x ticks=false,
xticklabel style={/pgf/number format/fixed},
]
\addplot[only marks, red, mark=o] table [x index={0}, y index={2}, col sep=semicolon]{data/file_avg.csv};\addlegendentry{Extension};
\addplot[only marks, blue, mark=+] table [x index={0}, y index={1}, col sep=semicolon]{data/file_avg.csv};\addlegendentry{NBLyzer};
\end{axis}
\end{tikzpicture}
\caption{Average Execution Run-times Per Notebook of Extension vs \textsc{NBLyzer}\label{fig:rt} }
\end{figure*}

We compare the runtimes of our analyzer (Extension) to the NBLyzer data leakage analysis from~\cite{NBLyzer} (NBLyzer) and show their average executions per notebook in Figure~\ref{fig:rt} (note, a single notebook may have several executions). Overall, our analysis experiences a slowdown of 7\%. For the vast majority, the difference in runtime is negligible. However, for several cases we experience a noticeable slowdown which we believe is due to our extended semantics that requires more data to be stored in the abstract state and more complex operations (join etc.). On the other hand, we also see the opposite occurring on a small number of benchmarks, where our analysis is faster. Here our speedup is due to discovering a bug and terminating execution while \textsc{NBLyzer} continues to propagate. As with \textsc{NBLyzer} the majority (over 99\%) of analyses completed in less than $1$ second except for a few cases. Pathological cases occur typically due to a notebook having large inter-connectivity between cells thus resulting in a large number of execution traces. 

Overall, we find that considering the increase in bug detection, the 7\% slowdown is a small price to pay and does not appear to significantly degrade the user experience for the interactive notebook use case. 

